\documentclass[10pt,twocolumn,showpacs,preprintnumbers,amsmath,amssymb,prb]{revtex4-1}
\usepackage{graphicx}  
\usepackage{dcolumn}   
\usepackage{bm}        
\usepackage[utf8]{inputenc}
\usepackage{graphics}
\usepackage{latexsym}
\usepackage{amsmath}
\usepackage{float}
\usepackage{cancel}
\usepackage{multirow} 
\usepackage{pstricks,pst-plot}
\newcommand{\bra}[1]{\ensuremath{\langle #1 |}}
\newcommand{\ket}[1]{\ensuremath{| #1 \rangle}}
\newcommand{\bracket}[2]{\ensuremath{\langle #1 | #2 \rangle}}

\begin{document}


\title{Orbital magnetic susceptibility of graphene and $\mathrm{MoS_2}$}
\author{A. Guti\'errez-Rubio,$^{1}$ T. Stauber,$^{1}$ G. G\'omez-Santos,$^{2}$ R. Asgari$^{3}$ and F. Guinea$^{4}$}
\affiliation{$^{1}$Departamento de Teor\'{\i}a y Simulaci\'on de Materiales, Instituto de Ciencia de Materiales de Madrid, CSIC, E-28049 Madrid, Spain}
\affiliation{$^2$ Departamento de F\'{\i}sica de la Materia Condensada, Instituto Nicol\'as Cabrera and Condensed Matter Physics Center (IFIMAC), Universidad Aut\'onoma de Madrid, E-28049 Madrid, Spain}
\affiliation{$^{3}$ School of Physics, Institute for Research in Fundamental Sciences (IPM), Tehran, 19395-5531, Iran}
\affiliation{$^{4}$ IMDEA Nanoscience Institute, E-28049 Madrid, Spain}
\date{\today}

\begin{abstract}
We calculate the orbital magnetic susceptibility $\chi_{\mathrm{orb}}$ for an 8-band tight-binding model of gapless and gapped graphene using Green's functions.
Analogously, we study $\chi_{\mathrm{orb}}$ for a $\mathrm{MoS_2}$ 12-band model. 
For both materials, we unravel the character of the processes involved in the magnetic response by looking at the contribution at each point of the Brillouin zone.
By this, a clear distinction between intra- and interband excitations is generally possible and we are able to predict qualitative features of $\chi_{\mathrm{orb}}$ only through the knowledge of the band structure.
The study is complemented by comparing the magnetic response with that of 2-band lattice Hamiltonians which reduce to the Dirac and Bernevig-Hughes-Zhang (BHZ) models in the continuum limit.
\end{abstract}
\pacs{.}
\maketitle


\section{Introduction}

Orbital magnetization in solids has gained  renewed attention in view of new two-dimensional materials with topologically non-trivial band structures.\cite{NovoselovPNAS05} Usually, this issue is addressed from either the perspective of isolated atoms or the picture of electron gases with a certain effective mass.\cite{Blount62,Misra69,ashcroft1976solid}  But semi classical approaches including geometrical effects due to a non-trivial Berry curvature\cite{Shi07,Xiao10,thonhauser2011theory,Gao14,raoux2015orbital,Gao15} or Green's function techniques\cite{fukuyama1971theory,Safran79,Koshino07,gomez2011measurable} offer new perspectives and thus reinforce the motivation for the present study.

Many developments, like the generalization of Fukuyama's formula to tight-binding systems,\cite{gomez2011measurable} are fairly recent.
Among the new phenomena arising from this approach, one can highlight the prediction of paramagnetism resulting from the periodic lattice potential due to a sum rule.
A paramagnetic orbital response also  occurs necessarily around van Hove singularities\cite{vignale1991orbital} and in Dirac systems the sublattice isospin degree of freedom gives rise to a contribution that can be interpreted as the traditional Pauli paramagnetism.\cite{koshino2010anomalous}
Moreover, contrarily to previous approaches via the Peierls-Landau formula\cite{peierls1933theory} and its generalization to multi-band systems, interband (or better geometrical) processes turn out to play a crucial role, e.g., filled bands need not be magnetically inert.\cite{Ominato13,raoux2015orbital}
In this context, it was shown that the band structure does not allow to uniquely determine the magnetic response of a solid, in stark contrast with the Peierls-Landau approach, i.e., different systems with an identical band structure can display completely opposite orbital magnetic responses.\cite{raoux2014dia}
The topological aspects of the band structure partly encoded in the Berry curvature play an important role in this scenario.\cite{thonhauser2011theory,raoux2015orbital}
In fact, using the semi classical wave-packet approach, a complete discussion about the several contributions to the magnetic susceptibility including purely geometrical terms was recently presented in Ref.
\onlinecite{Gao15}.

Our work aims at two prominent 2D materials which display a non-trivial topological band structure, namely graphene and $\mathrm{MoS_2}$.
Graphene is characterized by a Berry phase of $\pi$ manifested in the half-integer quantum Hall effect,\cite{Novoselov05,Zhang05} gapped graphene shows topological currents at zero magnetic field,\cite{Gorbachev14} and $\mathrm{MoS_2}$ is a topological valley insulator.\cite{Feng12,rostami2015valley}
It is thus worthwhile to count on a detailed characterization of their magnetic response through the discussion of the magnetic susceptibility.

In this article, we will perform the numerical calculation of the orbital magnetic susceptibility, $\chi_{\mathrm{orb}}$, for multi-band tight-binding models using the Green's function formalism.
For graphene, we will deal with a nearest-neighbor 8-band model including all $2s$ and $2p$ orbitals.
To model $\mathrm{MoS_2}$, the relevant bands are formed by $d$-orbitals with a small influence of $p$-orbitals amounting to an effective 12-band model.\cite{Kormanyos13,Cappelluti13,rostami2015valley} 

We also discuss the nature of the processes involved in the magnetic response by analyzing the contribution of each point of the first Brillouin zone to $\chi_{\mathrm{orb}}$.
The action of processes related to the Fermi surface or to geometrical effects can be distinguished by means of this approach, which yields valuable information to physically understand the magnetic response of solids. 
Finally, we address the magnetic response by means of several 2-band tight-binding as well as continuum models.
The latter are mostly valid for energies close to the valence and conduction bands, but also suit other parts of the spectrum like the Dirac gap in between the second and third core bands of $\mathrm{MoS_2}$.
We analyze the sources underlying all these facts, the Berry curvature playing a crucial role, and discuss the seek for a 2-band model that yields an accurate continuum description of $\mathrm{MoS_2}$ at the neutrality point.
This will allow us to address a still debated question about the magnetic response: under which conditions $\chi_{\mathrm{orb}}$ can be qualitatively extracted from the mere knowledge of the band structure.

The paper is organized as follows.
The formalism for calculating the orbital magnetic susceptibility of a general tight-binding model as well as previous approaches are recalled in Sec. \ref{sec_tight-binding} and in an appendix.
In Sec. \ref{sec_chigraphene}, we introduce the Hamiltonians used to describe gapless and gapped graphene, calculate and compare their respective magnetic susceptibilities, and relate these to the features of the band structure.
Sec. \ref{sec_chimos2} is analogous but devoted to $\mathrm{MoS_2}$.
Sec. \ref{sec_effective} establishes a comparison with effective models for both materials, and finally our conclusions are presented in Sec. \ref{sec_conclusion}.

\section{Magnetic susceptibility of tight-binding models}
\label{sec_tight-binding}
We numerically calculate the orbital magnetic susceptibility using the following formula, valid for arbitrary tight-binding models:\cite{gomez2011measurable}
\begin{align}
\chi_{\mathrm{orb}}&=
-\frac{\mu_0e^2}{2\pi\hbar^2}
\times
\notag
\mathrm{Im}\int_{-\infty}^{\infty}dE\,n_F(E)\\
&\times\;\frac{1}{A}\sum_{\vec{k}}\mathrm{Tr}
\Bigg\{
\hat{\gamma}^x\hat{G}\hat{\gamma}^y\hat{G}\hat{\gamma}^x\hat{G}\hat{\gamma}^y\hat{G}
+
\notag
\\
&
+\frac{1}{2}(\hat{G}\hat{\gamma}^x\hat{G}\hat{\gamma}^y+\hat{G}\hat{\gamma}^y\hat{G}\hat{\gamma}^x)\hat{G}\frac{\partial\hat{\gamma}^y}{\partial k_x}
\Bigg\}
,
\label{chiorb}
\end{align}
where $\hat{G} = (E-H_{\vec{k}}+i0^+)^{-1}$, $\hat{\gamma}^{x,y}=\partial H_{\vec{k}}/\partial k_{x,y}$ and $H_{\vec{k}}$ is the Hamiltonian at wave vector $\vec{k}$ including the spin degree of freedom.
Further, $\mu_0$ is the vacuum permeability, $A$ denotes the sample area, and $n_F(E)=(e^{(E-\mu)/T}+1)^{-1}$ is the Fermi function.
In the following, we will present results at $T=0$ with $\mu=E_F$ the Fermi energy.

To derive the gauge-invariant magnetic susceptibility for a general tight-binding model, the correct wave vector-dependence of the current operator needs to be used.\cite{stauber2010dynamical} Only then the gauge-dependent contribution of the diamagnetic current is canceled, see appendix.
The longitudinal response can be obtained from the above formula by replacing the $y$-superindexes with $x$ and vice versa.
It must necessarily be zero due to gauge invariance which is guaranteed by the exact cancellation of the first term by the second term.
Let us also highlight the sum rule\cite{gomez2011measurable} 
\begin{align}
\label{SumRule}
\int_{-\infty}^{\infty}dE_F\;\chi_{\mathrm{orb}}(E_F)=0\;,
\end{align}
which is obtained from the fact that $\chi_{\mathrm{orb}}(E_F)$ can be analytically continued into the upper complex plane, together with the residuum theorem.
Details on the above discussion can be found in the appendix.

\subsection{Previous approaches}
We recall that the first term in brackets of Eq. (\ref{chiorb}) yields the Fukuyama formula,\cite{fukuyama1971theory} which is valid for a Galilean invariant system with a possible linear term in $\vec{k}$, i.e., for all models with $\frac{\partial\hat{\gamma}^y}{\partial k_x}=0$.
But also for isotropic models like the tight-binding model for graphene involving only the $p_z$-orbitals, this term is dominant and the second term can almost be neglected.
However, in the case of the tight-binding models for graphene involving $s$-orbitals, the second term becomes quantitatively important, i.e., $\frac{\partial\hat{\gamma}^y}{\partial k_x}$ is not small due to the directional $\sigma$-bonds.

An even earlier approach is given by the Peierls-Landau orbital susceptibility and its trivial extension to multi-band systems,\cite{peierls1933theory,fukuyama1971theory}
\begin{align}
\chi_{\mathrm{PL}} =
\frac{\mu_0e^2}{12\hbar^2}
\frac{1}{A}\sum_{\vec{k},n}n'_F(\epsilon_{n,\vec{k}})
\left[
\epsilon^{xx}_{n,\vec{k}}\epsilon^{yy}_{n,\vec{k}} - (\epsilon^{xy}_{n,\vec{k}})^2
\right].
\label{chiPL}
\end{align}
Here, $n'_F(\epsilon)$ is the derivative of the Fermi-Dirac distribution, $\epsilon_{n,\vec{k}}^{x_ix_j}$ denotes the derivative of the energy eigenvalues with respect to $k_{x_i}$ and $k_{x_j}$, and $n$ is the band index.

Remarkably, $\chi_{\mathrm{PL}}$ only depends on the dispersion relation, whereas in Eq. (\ref{chiorb}), further information concerning the features of the eigenstates is contained.
Furthermore, due to $n'_F$, only states around the Fermi surface contribute to Eq. (\ref{chiPL}).
This is again in contrast with Eq. (\ref{chiorb}), where matrix instead of scalar multiplications properly include all contributions originating from possible interband transitions.
These differences turn out to be crucial in the appropriate description of the magnetic response of a multi-band tight-binding system, as we will show throughout this work.

\subsection{Continuum models and lattice contribution}
Several prominent features of the magnetic response of systems with a direct band gap at the $K$-points can be understood from the effective continuum model of gapped Dirac fermions:
\begin{align}
\chi_{\mathrm{Dirac}}
&=-\frac{g_sg_v\mu_0e^2}{6\pi}\frac{1}{2m_{\mathrm{eff}}}\Theta(|\Delta|-|E_F|),
\label{chiDiracgap}
\end{align}
where $g_s$ and $g_v$ are the spin and valley degeneracy, respectively, $2|\Delta|$ is the gap and $m_{\mathrm{eff}}$ is the mean effective mass derived from the curvature of the bands close to it.
For a Dirac model with constant gap, we have $m_{\mathrm{eff}}=|\Delta|/v_F^2$, with $v_F$ the Fermi velocity.
For chemical potentials inside the gap, $\chi_{\mathrm{Dirac}}$ is equal to the geometrical susceptibility as introduced in Ref. \onlinecite{Gao15}.

For a general gap with a $k^2$-dependence ($\Delta\to\Delta+\beta k^2$), this is modified to $m_{\mathrm{eff}}=[(\hbar v_F)^2+\beta\Delta]/(\hbar^2|\Delta|)$.
In the limit of $\Delta\beta\to0$, the step function becomes a Dirac delta leading to the following expression first obtained by McClure:\cite{McClure56,McClure60}
\begin{align}
\chi_{\mathrm{Dirac}}
&=-\frac{g_sg_v\mu_0e^2v_F^2}{6\pi}\delta(E_F-E_{DC}) 
\label{chiDiracgapless}
\end{align}
We included the constant energy shift $E_{DC}=0$ to indicate the location of the Dirac cone, needed for subsequent generalizations.
In fact, the above formulas also hold for $H=\sum_{\vec k} \hbar(\vec v_1\cdot\vec k)\sigma_x+\hbar(\vec v_2\cdot\vec k)\sigma_y+\Delta\sigma_z$ by replacing $v_F^2\to|\vec v_1\times\vec v_2|_z$, so more general band-crossings ($\Delta=0$) or gaps ($\Delta\neq0$) display the same features of the response.

If we describe graphene by a single orbital tight-binding model with only nearest-neighbor hoppings, lattice effects can be separated from the contribution that come from the continuum model.
We can then define the lattice susceptibility as\cite{gomez2011measurable} 
\begin{align}
\chi_{\mathrm{lattice}}\equiv\chi_{\mathrm{orb}}-\chi_{\mathrm{Dirac}} 
\end{align}
valid for the gapless or gapped case.
Using the following unit of the susceptibility
\begin{align*}
\chi_0=\frac{\mu_0e^2|t|a^2}{\hbar^2},
\end{align*}
we find that $\chi_{\mathrm{lattice}}/\chi_0$ is now scale-invariant, i.e., independent of the parameters of the model, where $t$ is the hopping amplitude and $a$ the lattice constant.
In the continuum limit $a\to0$ keeping $\frac{3at}{2\hbar}=v_F=\mathrm{const.}$, $\chi_0\to0$ and the lattice contribution thus also tends to zero. 

In the case of multi-band Hamiltonians with several hopping parameters, such a simple, scale-invariant quantity cannot be defined, especially not in the case of $\mathrm{MoS_2}$.
Still, we will present all results in units of $\chi_0$ and use $t=2.8\,\mathrm{eV}$ and $t=1.6\,\mathrm{eV}$ for graphene and $\mathrm{MoS_2}$, respectively.
Notice that then $\chi_0/a\sim\alpha\frac{v_F}{c}$ is the natural scale of the magnetic susceptibility with $\alpha\approx1/137$ the fine-structure constant.
 
In the subsequent sections, we will use the above definitions and proceed to apply these expressions to different tight-binding systems modeling graphene and $\mathrm{MoS_2}$.
Special attention will be paid to interpret the physics encoded in Eq. (\ref{chiorb}), to the relation of the results with the underlying band structure and to the possibility of finding effective models that yield a correct description of the magnetic response of these materials.

\section{Magnetic susceptibility of graphene}
\label{sec_chigraphene}

\begin{figure}
\includegraphics{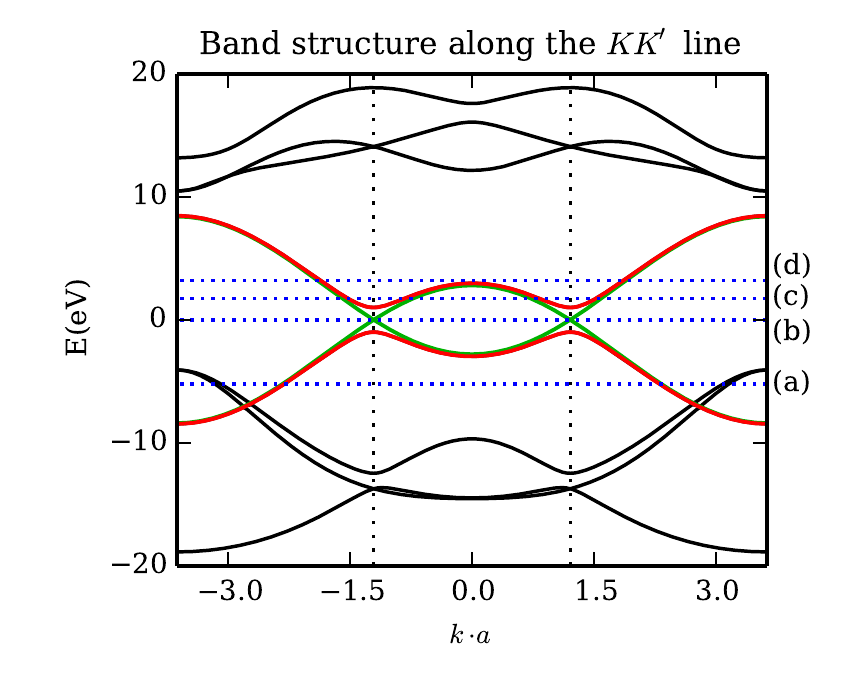}
\caption{
(Color online)
Band structure of the Slater-Koster model including the $\sigma$-bonds (black lines) and $\pi$-bonds (red and green referring to the gapless and gapped case, respectively).
Vertical dotted lines indicate the position of $K$ and $K'$ points.
Horizontal blue dotted lines labeled with a letter are respective to the Fermi energies of Fig. \ref{chikpanelgraphene}.
}
\label{bandstructureGSK}
\end{figure}

In this section, we discuss the magnetic response of graphene within the Slater-Koster description including all four orbitals of the valence band, i.e., $2s$, $2p_x$, $2p_y$, and $2p_z$.
The parameters of the hopping elements and energies are adopted from Ref. \onlinecite{shengjun2015electronic}.
 In Fig. \ref{bandstructureGSK}, the band structure is shown in the $KK'$ direction for the gapped ($\Delta=1\,\mathrm{eV}$) and gapless case. 

\begin{figure*}
\includegraphics{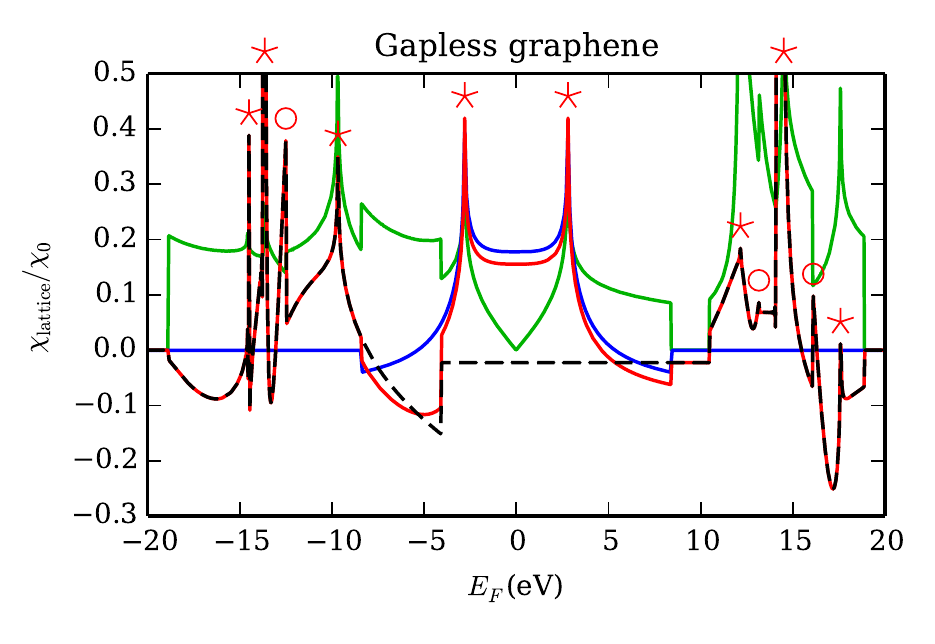}
\includegraphics{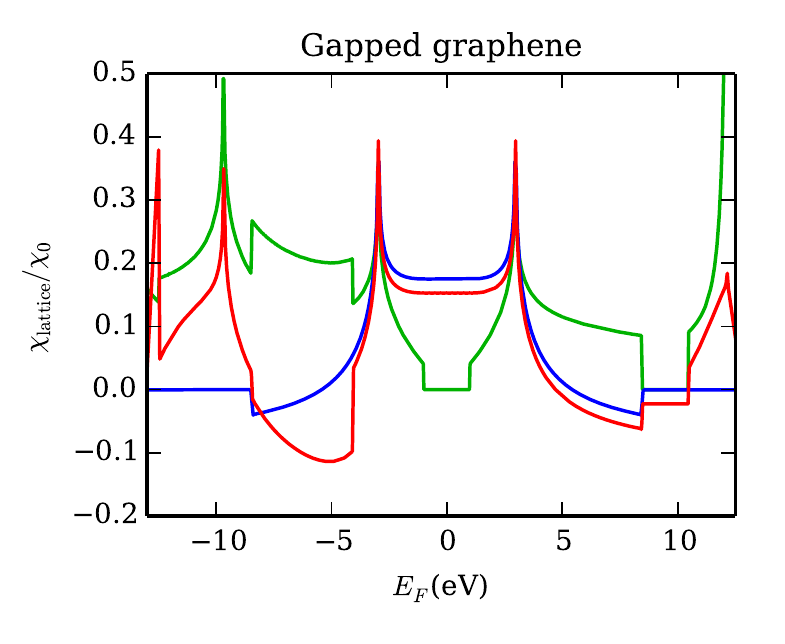}
\caption{
(Color online)
The lattice susceptibility $\chi_{\mathrm{lattice}}$ for the 8- (red) and 2-band (blue) models of graphene described in the main text.
For the gapped case, a closeup view is chosen for the sake of clarity.
Actually, the parts of the curves outside of the plot range coincide with those of the left figure.
The dashed black line depicts $\chi_{\mathrm{orb}}$ of the $\sigma$-bands.
The density of states is plotted in green.
In the left figure, the asterisks over the peaks denote that they arise due to van Hove singularities, whereas circles are placed over finite discontinuities coming from band edges.
}
\label{chiorbGSK}
\end{figure*} 
 
The lattice contribution of the magnetic orbital susceptibility for gapless graphene appears on the left hand side of Fig. \ref{chiorbGSK} as a red curve.
Since the $\sigma$-bands (black dashed curve) and $\pi$-bands (blue curve) decouple due to symmetry, the $\pi$-bands yield a contribution to $\chi_{\mathrm{orb}}$ identical to that discussed in Ref. \onlinecite{gomez2011measurable}.
Further, the total susceptibility and the one coming from the $\sigma$-bands coincide for low and high energies.
For comparison, also the density of states is shown as a green curve. 

For Fermi energies around half filling, the lattice contribution displays a constant plateau that evolves into the expected paramagnetic divergences when hitting the van Hove singularities.\cite{vignale1991orbital}
Due to the presence of at least one van Hove singularity in each band, $\chi_{\mathrm{orb}}$ shows a quite irregular structure.
We identify the resulting paramagnetic divergences in Fig. \ref{chiorbGSK}, making a distinction with respect to finite discontinuities that come from band edges.
Although the latter might be unraveled by peak asymmetry, our conclusions have been drawn from a careful analysis of the band structure.
A diamagnetic response is found for Fermi energies in the intervals $(-19,-15)\,\mathrm{eV}$ and $(-8,-4)\,\mathrm{eV}$, as is suggested by the parabolic dispersion relation of the corresponding bands, i.e., Landau diamagnetism.

The full magnetic orbital response shows two delta-like diamagnetic peaks at Fermi energies $E_F\approx\pm14$eV associated with the band-crossing at the $K$-points.
We were able to subtract these contributions using Eq. (\ref{chiDiracgapless}) with $v_F\simeq 3.9\cdot 10^5\,\mathrm{m/s}$ ($v_F\simeq 3.5\cdot 10^5\,\mathrm{m/s}$) for the lower (upper) crossing at $E_{DC}\approx\pm14$eV, where the Fermi velocities and Dirac cone energies were extracted from the band-structure.
In Fig. \ref{chiorbGSK}, we thus plot the generalized lattice contribution 
\begin{align}
\chi_{\mathrm{lattice}}\equiv\chi_{\mathrm{orb}}-\chi_{\mathrm{Dirac}}-\chi_{\mathrm{Dirac}}^\sigma\;,
\end{align}
where $\chi_{\mathrm{Dirac}}^\sigma$ denotes the above delta-like Dirac contributions.

Let us now comment on the constant diamagnetic contribution from the $\sigma$-bands inside the gap around the $\Gamma$-point which amounts to an extra $\sim 12\%$ lattice contribution to the $\pi$-electrons.
For gated graphene away from half-filling and at low temperatures, we thus expect a measurable contribution of the $\sigma$-bands to the magnetic susceptibility.
This response is of pure geometrical nature as we will argue below.

On the right hand side of Fig. \ref{chiorbGSK}, we show the total (red) and $\pi$-band (blue) lattice contribution $\chi_{\mathrm{lattice}}$ of the gapped graphene model with $\Delta=1\,\mathrm{eV}$ for energies around the neutrality point.
Interestingly, both models, gapless and gapped graphene, show an almost identical lattice contribution even at the energies where their spectra strongly differ, namely close to the neutrality point, as can be appreciated form the density of states (green curve) on the left and right side of Fig. \ref{chiorbGSK}.
This is an indicative of the fact that the Dirac model is the continuum version of the lattice models under consideration.

\subsection{Brillouin zone analysis}
In order to count on a deeper understanding of the above results, we proceed to discuss the individual contribution of each point of the first Brillouin zone to $\chi_{\mathrm{orb}}$, i.e., we plot $\bar{\chi}_{\mathrm{orb}}(\vec{k},E_F)$ for the first Brillouin zone with $\chi_{\mathrm{orb}}(E_F)=\sum_{\vec{k}}\bar{\chi}_{\mathrm{orb}}(\vec{k},E_F)$.
Our approach is intended to inquire about Eq. (\ref{chiorb}) in more detail and to unravel the physics behind it.
Last but not least, it will serve as a tool to compare the magnetic response of the different models considered, see Sec. V.

\begin{figure}
\includegraphics{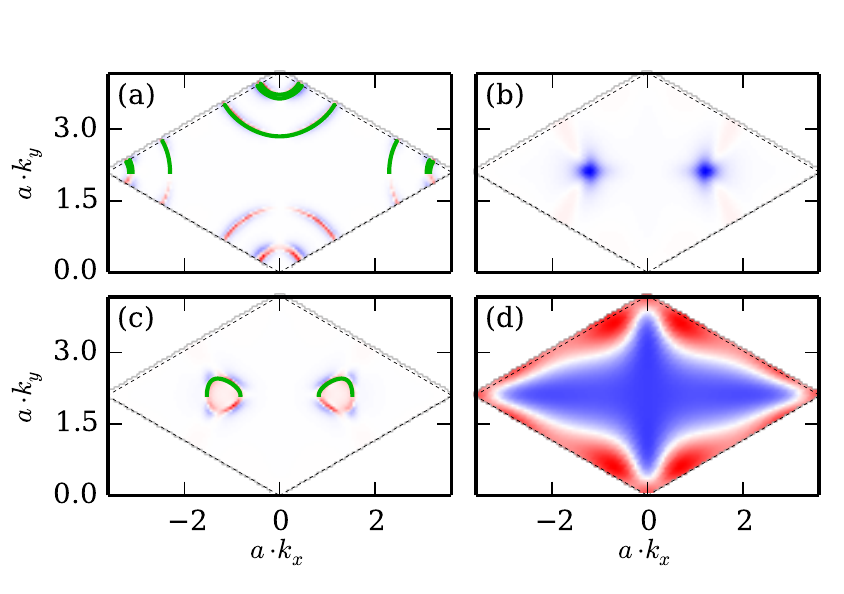}
\caption{
(Color online)
Contribution of each point of the Brillouin zone to $\chi_{\mathrm{orb}}$ for a fixed Fermi energy $E_F$, as defined in the main text. 
Paramagnetism and diamagnetism correspond to red and blue, respectively, the color scale being normalized to $\max|\bar{\chi}_{\mathrm{orb}}(\vec{k},E_F)|$ for the given $E_F$.
The green lines depict the Fermi surface.
For the sake of clarity, they appear only in the upper half of each plot, but can be extended to the lower one by an horizontal mirror reflection.
$E_F$ takes the values $-5.2$ (a), $0$ (b), $1.75$ (c), and $3.2\,\mathrm{eV}$ (d), which are indicated in Fig. \ref{bandstructureGSK} and labeled there with the corresponding letter.
In the last case, only the contribution of the $\sigma$-bands has been considered.
An imaginary part of the energy equal to $0.3\,\mathrm{eV}$ has been used in these calculations.
}
\label{chikpanelgraphene}
\end{figure}

In the following, we will focus on the case of gapped graphene for simplicity.
The results for the first Brillouin zone are plotted in Fig. \ref{chikpanelgraphene}.
It can be seen that the contributions to $\chi_{\mathrm{orb}}$ mostly come from points of the Brillouin zone that are pinned to the Fermi surface; the other states practically remain inert.
Figs. \ref{chikpanelgraphene} (a) and (c) depict this situation, dealing with more complex or simpler regions of the band structure, respectively.
We also note that the response can be either diamagnetic (blue) or paramagnetic (red).

Let us now discuss the situation where the Fermi energy lies inside a gap.
In fact, the highest diamagnetism is found for $E_F$ inside the Dirac-like gap in clear contrast to the predictions of the Peierls-Landau formula.
The $\vec{k}$-points contributing to the susceptibility are now concentrated in small regions around $K$ and $K'$ points as seen in Fig. \ref{chikpanelgraphene} (b).
Fig. \ref{chikpanelgraphene} (d), on the other hand, addresses the origin of the constant diamagnetic plateau coming from $\sigma$-bands close to neutrality (cf. the dashed black line in Fig. \ref{chiorbGSK}).
Interestingly, the magnetic response is now smeared throughout the whole Brillouin zone rather than being concentrated, e.g., around the $\Gamma$-point.  

We conclude that in principle there exist clear mechanisms underlying the magnetic response of a tight-binding system and they are strongly related to the band structure.
For Dirac gaps around the $K$-points, transitions only around these points are relevant, whereas for gaps around the $\Gamma$-point, transitions in the whole Brillouin zone contribute to the final response.

In the next section, we will extend our analysis to a more complex system, i.e., transition metal dichalcogenides in form of  $\mathrm{MoS_2}$.

\section{Magnetic susceptibility of $\mathrm{MoS_2}$}
\label{sec_chimos2}

\begin{figure}
\begin{center}
\includegraphics{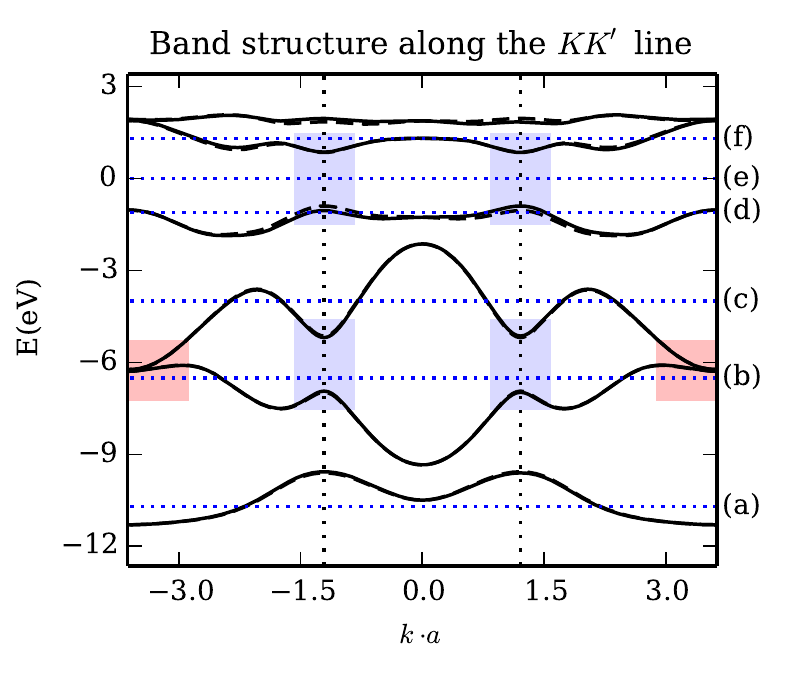}
\end{center}
\caption{
(Color online)
Section of the band structure of the $\mathrm{MoS_2}$ 12-band model of Ref. \onlinecite{rostami2015valley}.
Solid (dashed) lines correspond to spin $s=(-)1$.
The gaps and band overlap discussed in the main text have been highlighted in light blue and red, respectively.
A close-up view of the valence and conduction bands appears in Fig. \ref{FigBHZ} (B).
Vertical dotted lines indicate the position of $K$ and $K'$ points.
Horizontal blue dotted lines labeled with a letter are respective to the Fermi energies of Fig. \ref{chikpanelMoS2}.
}
\label{bandstructure}
\end{figure}

In this section, we discuss the orbital susceptibility for the 12-band Hamiltonian derived in Refs. \onlinecite{Cappelluti13,Kormanyos13,rostami2015valley}.
A plot of the band structure along the $K-K'$ direction is shown in Fig. \ref{bandstructure}.
Let us point out that there are two Dirac-like gaps centered at $K$ and $K'$: one between the second and third bands and another one between the valence and conduction bands.

\begin{figure*}
\begin{center}
\includegraphics{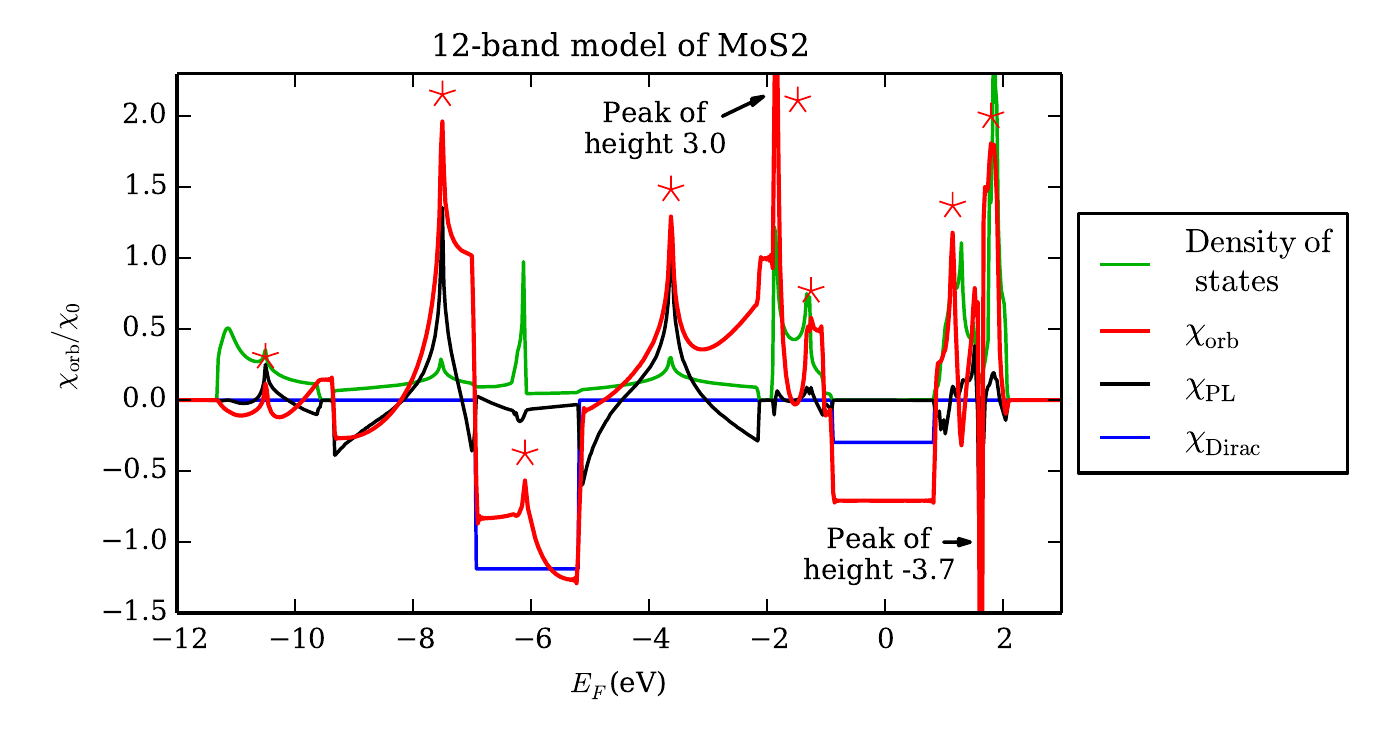}
\end{center}
\caption{(Color online) Magnetic orbital susceptibility of the $\mathrm{MoS_2}$ 12-band model using Eq. (\ref{chiorb}) (red) and Eq. (\ref{chiPL}) (black) and density of states (green).
The text inside the plot corresponds to the red curve.
Also shown the Dirac susceptibility of Eq. (\ref{chiDiracgap}) for the gaps highlighted in Fig. \ref{bandstructure} (blue).
The asterisks mark the peaks associated to van Hove singularities.
}
\label{chiMoS2}
\end{figure*}

The full magnetic orbital susceptibility $\chi_{\mathrm{orb}}$ as calculated by means of Eq. (\ref{chiorb}) is plotted in Fig. \ref{chiMoS2} as a red curve.
We compare the results with the magnetic orbital response of the Peierls-Landau formula, Eq. (\ref{chiPL}), seen as a black line, together with the density of states (green line).
Let us also comment on the two diamagnetic regions at Fermi energies matching those of the aforementioned gaps, which are highlighted (blue) in Fig. \ref{bandstructure}.
Their expected magnetic susceptibility according to the Dirac continuum model, Eq. (\ref{chiDiracgap}), is shown in Fig. \ref{chiMoS2} as a blue line.


\begin{figure*}
\includegraphics{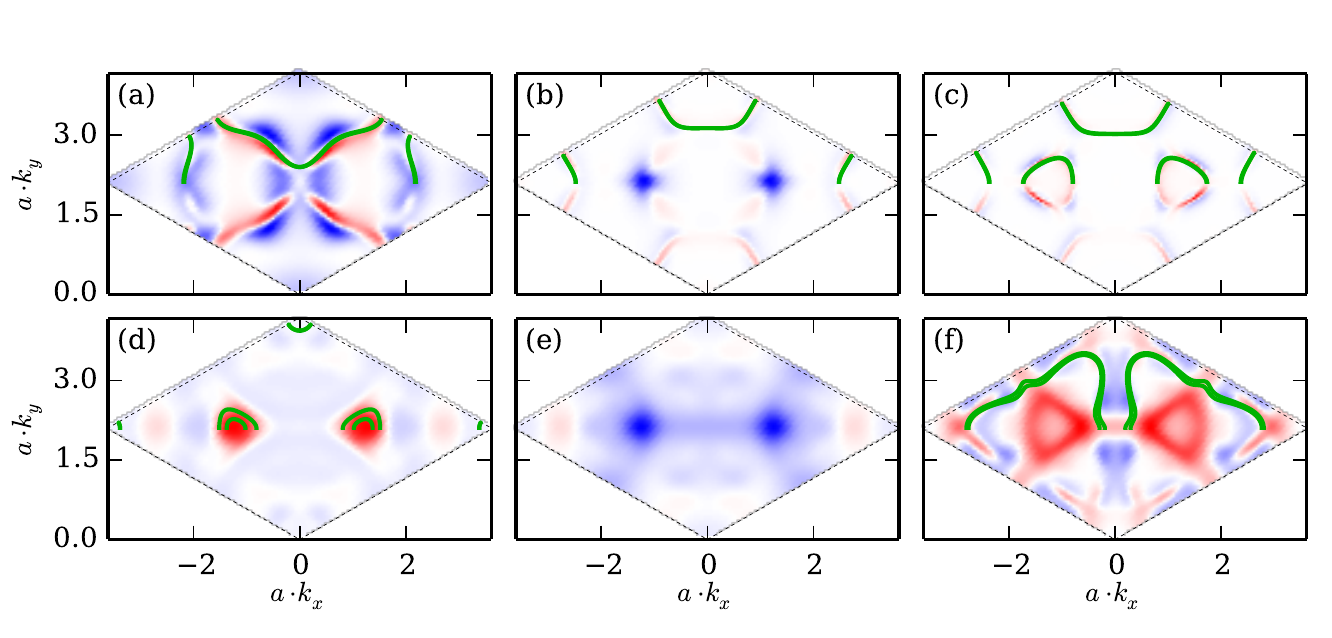}
\caption{
(Color online)
Same as Fig. \ref{chikpanelgraphene} but for the 12-band model of $\mathrm{MoS_2}$.
$E_F$ takes the values $-10.7$ (a), $-6.5$ (b), $-4$ (c), $-1.1$ (d), $0$ (e) and $1.3\,\mathrm{eV}$ (f), which are indicated in Fig. \ref{bandstructure} and labeled there with the corresponding letter.
An imaginary part of the energy equal to $0.3\,\mathrm{eV}$ has been used in these calculations.
}
\label{chikpanelMoS2}
\end{figure*}

\begin{figure*}
\begin{center}
\includegraphics{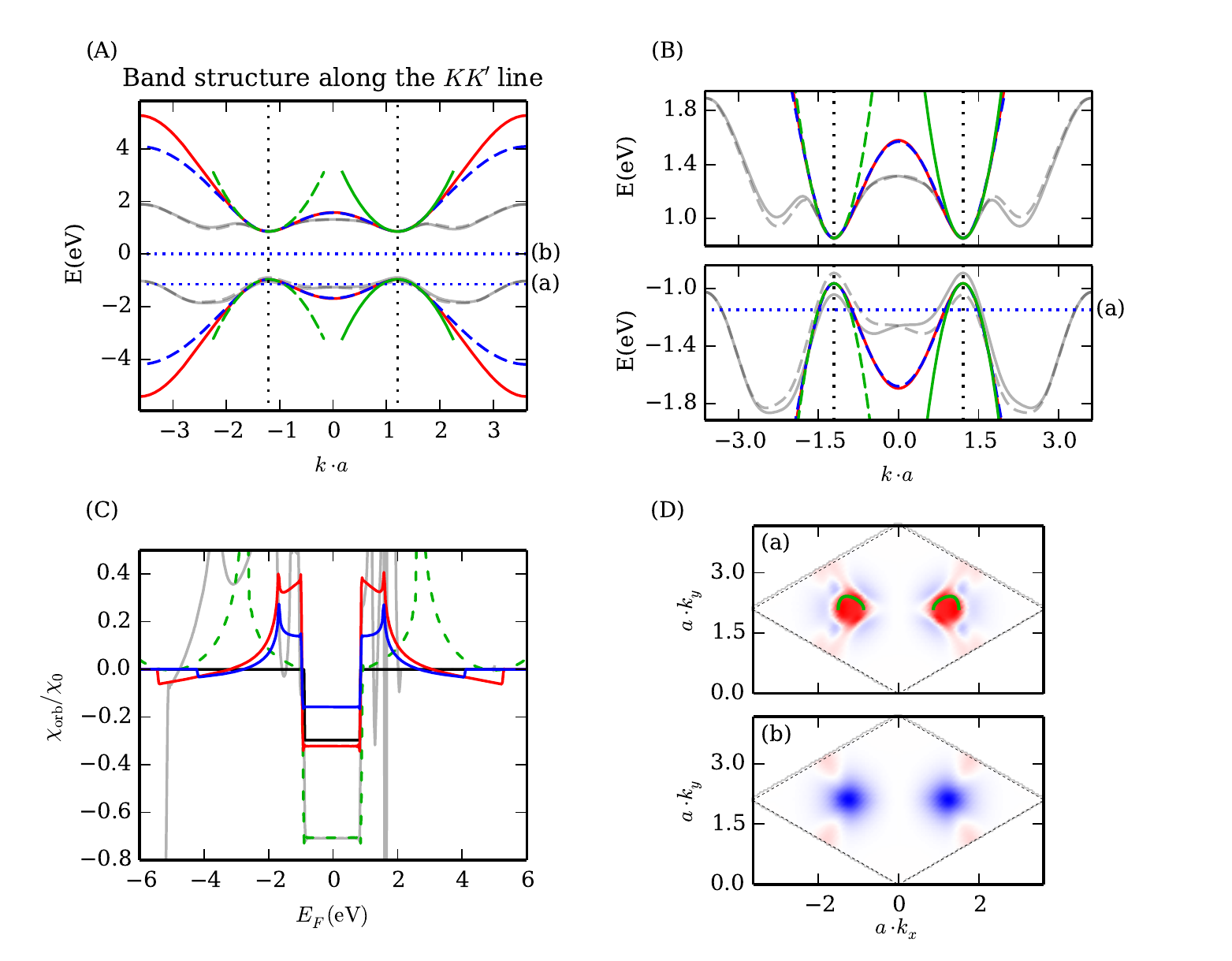}
\end{center}
\caption{
(Color online)
Band structure (A) and close-up view (B) for the 2-band lattice models of Eqs. (\ref{HSQ}) (green) and (\ref{HBHZ}) with parameters given by Eqs. (\ref{choice1}) (red) and Eqs. (\ref{choice2}) (dashed blue).
Gray corresponds to the valence and conduction bands of $\mathrm{MoS}_2$.
Due to the differences in the unit cells, only the parabolic approximation ---close to $k=0$--- of the square lattice bands is plotted near $K$ (solid) and $K'$ (dashed lines) points.
Vertical dotted lines indicate the position of $K$ and $K'$ points, and horizontal blue dotted lines labeled with a letter are respective to the Fermi energies of (D).
(C) depicts the magnetic susceptibilities with the same color code as (A) and (B).
The extra black line depicts $\chi_{\mathrm{Dirac}}/\chi_0$.
(D) shows two subplots analogous to those of Fig. \ref{chikpanelgraphene} but for the model of Eqs. (\ref{choice1}) and (\ref{HBHZ}).
The corresponding Fermi energies are $-1.15\,\mathrm{eV}$ (a) and $0$ (b).
}
\label{FigBHZ}
\end{figure*}

Apart from diamagnetic regions associated to Dirac gaps or parabolic bands, we have again identified the paramagnetic peaks corresponding to van Hove singularities.
Some of them are also reproduced by the Peierls-Landau magnetic susceptibility.
The latter fails to yield other relevant features, though, above all concerning the magnetic response of filled bands.
Let us also comment on the diamagnetic peak at $E_F\approx1.7$eV with height $-3.7\chi_0$.
This peak can be associated to four gapped Dirac cones located at close vicinities of the $\Gamma \mathrm{M}$ direction, for which the Berry curvature yields large values.
Interesting physics might be expected to emerge from them, especially regarding their spin split character and high directional asymmetry.
As for the experimental realization, reaching the corresponding Fermi energies could be overcome in the future by the use of liquid dielectric capacitors.

The red curve of Fig. \ref{chiMoS2} depicts one of the main results of this study, which can be experimentally verified, especially for neutral $\mathrm{MoS_2}$ with the Fermi energy inside the gap.
We will now continue with the Brillouin zone analysis for $\mathrm{MoS_2}$.

\subsection{Brillouin zone analysis}
As in Sec. \ref{sec_chigraphene}, we proceed to study the individual contribution of each point inside the 1st Brilluoin zone to $\chi_{\mathrm{orb}}$.
Fig. \ref{chikpanelMoS2} displays the results for this analysis, on the one hand confirming the conclusions we extracted from graphene and on the other hand offering more information to be discussed.

In the case of $\mathrm{MoS_2}$, the subplots are now more diverse and complex than in the case of graphene.
Still, we can associate most features to processes involving the Fermi surface, such as subplots (a), (c), and (d).
In the case of subplot (b), we see contributions from the Fermi surface located around the $\Gamma$-point as well as processes involving the two $K$-points.
This has been discussed above and the diamagnetic contribution due to the Dirac-like gap is shown as blue line in Fig. \ref{chiMoS2}.
Also for Fermi energies inside the valence and conduction band, the main contribution comes from the two $K$-points as can be seen in subplot (e).
However, the structure is considerably more involved, indicating that the simple Dirac model does not quantitatively reproduce the diamagnetic response.

Let us finally discuss the spectrum shown in subplot (f), displaying features that were not seen in the case of graphene.
Again, there is a contribution associated to the Fermi surface.
But we also see a paramagnetic response displaying a prominent trigonal shape reminiscent of the Fermi surface of graphene at the $M$-point.

\section{Effective models}
\label{sec_effective}
We will now analyze reduced, effective models and test them with respect to their predictive reliability.
We will first summarize the results for the continuum Dirac model and then look at two-band lattice models.
Fig. \ref{FigBHZ} gathers all the results.

\subsection{Effective continuum model}
\label{EffectiveCM}
Let us summarize to what extent the continuum model of Eq. (\ref{chiDiracgap}) agrees with the results yielded by Eq. (\ref{chiorb}).
For graphene, we already concluded that the continuum model yields the main contribution to the magnetic response around the neutrality point.
This is best seen from the lattice contribution $\chi_{\mathrm{lattice}}$ close to half filling, which shows a constant paramagnetic offset independent of the actual gap coming from the orbital response of electrons outside the Dirac cone region.
Also for energies at the two Dirac cones of the $\sigma$-band, the main contribution is given by Eq. (\ref{chiDiracgapless}).

For $\mathrm{MoS_2}$, the discussion is more subtle  and different for the two gaps present in the band structure.
For the core gap, the diamagnetic depth is fairly reproduced by the Dirac well.
The finer structure of the curve that corresponds to the full spectrum can be explained as a consequence of the band overlap at the corners of the 1st Brillouin zone, cf. Fig. \ref{bandstructure}, and the presence of a van Hove singularity within that energy range.
On the other hand, when considering the gap between the valence and conduction band, there is only a qualitative agreement with the result of the Dirac continuous model, i.e., the well-like response is correctly reproduced, but the numerical value is off by approximately a factor of two. In fact, Fig. \ref{chikpanelMoS2}(e) indicates that the gap is not well reproduced by assuming a simple Dirac-like gap (with and without a $k^2$-dependent mass).
Also the product between the Berry curvature and the magnetic orbital moment as suggested by Ref. \onlinecite{Gao15} do not yield contributions isotropically concentrated around the $K$-points.
An effective continuum model to correctly describe the trigonal feature of the magnetic response and Berry curvature is thus still uncovered.

\subsection{Effective two-band lattice models}
For the magnetic orbital susceptibility of $\mathrm{MoS_2}$ around the neutrality point, we carry out a comparison with three different two-band lattice models to inquire about possible fits.
They consist of a two-orbital square lattice and of a one-orbital hexagonal lattice with and without next-nearest neighbor coupling.
We impose the constraint that they reproduce the energies of the 2-band effective Hamiltonian of Ref. \onlinecite{rostami2015valley}, although neglecting the spin-valley splitting, see the close-up of the band structure in Fig. \ref{FigBHZ} (B).
The hallmark of this effective description is the $k^2$-dependence of the potential and mass terms.

Firstly, let us introduce the tight-binding Hamiltonian defined on  a square lattice with two orbitals at each site.
The off-diagonal terms have their origin in a spin-orbit coupling.
We thus have
\begin{align}
    H_{\vec{k}} =
        \left(
            \begin{array}{cc}
                \frac{\Delta_0+\Delta}{2} + 2 b \beta c_{\vec{k}} 		&
                -t_0	s_{\vec{k}}										\\
                -t_0s_{\vec{k}}^*										&
                \frac{\Delta_0-\Delta}{2} - 2 b \beta c_{\vec{k}}
            \end{array}
        \right)\;,
    \label{HSQ}
\end{align}
with $c_{\vec{k}}=2+\cos(ak_x) + \cos(ak_y)$ and $s_{\vec{k}}=\sin(ak_x) - i \sin(ak_y)$.
We use the following parameters in order to match the band structure of the $\mathrm{MoS_2}$ 12-band model. 
\begin{align}
&
\Delta_0=-0.11\,\mathrm{eV},\ 
\Delta=1.82\,\mathrm{eV},\ 
t_0=2.33\,\mathrm{eV},\notag
\\
&
\alpha=-0.01,\ 
\beta=-1.54
\label{choice1}
\end{align}
and $b=\hbar^2/(4m_0a^2)\simeq 0.572\;\mathrm{eV}$.
Note that the contribution of $\alpha$ is neglected in the above Hamiltonian, but will be included below.

Secondly, the continuum model with quadratic mass and scalar potential can be deduced from a hexagonal lattice with a single orbital per site and next-nearest neighbor hoppings.
Choosing the hopping parameters and on-site energies accordingly, we arrive at
\begin{align}
H_{\vec{k}}=
\left(
\begin{array}{cc}
\frac{\Delta_0+\Delta}{2}+\frac{4b(\alpha+\beta)}{9}|\phi_{\vec{k}}|^2
&
-\frac{2}{3}t_0\phi_{\vec{k}}
\\
-\frac{2}{3}t_0\phi^*_{\vec{k}}
&
\frac{\Delta_0-\Delta}{2}+\frac{4b(\alpha-\beta)}{9}|\phi_{\vec{k}}|^2
\end{array}
\right),
\label{HBHZ}
\end{align}
with the form factor $\phi_{\vec{k}}=\sum_{j}e^{i\vec{\delta}_j\cdot\vec{k}}$ and $\vec{\delta}_j$ ($j=1,2,3$) the three vectors joining nearest neighbors.

At last, the gapped Dirac lattice model with constant mass can be easily reproduced with a different choice of the parameters in Eq. (\ref{HBHZ}):
\begin{align}
&
\Delta_0=-0.11\,\mathrm{eV},\ 
\Delta=1.82\,\mathrm{eV},\ 
t_0=2.02\,\mathrm{eV},\notag
\\
&
\alpha=\beta=0.
\label{choice2}
\end{align}

All models display the same curvature (mass) around the $K$-points as can be seen from the band structures which are plotted over part of the $\mathrm{MoS_2}$ spectrum in Figs. \ref{FigBHZ} (A) and (B).

\subsection{Discussion}
Here, we will compare the magnetic response of the effective two-band models with that of Sec. \ref{sec_chimos2}.
To do so, we will include the necessary spin and valley degeneracy factors $g_s$ and $g_v$, respectively, i.e., for the square lattice we include a factor $g_sg_v$ whereas for the hexagonal model only a factor $g_s$.

From Fig. \ref{FigBHZ} (C), one can appreciate a significant quantitative difference between the different models even for Fermi energies inside the gap.
This fact points at the discussion of Refs. \onlinecite{raoux2014dia,raoux2015orbital}, namely that a mere match of the band structure of two different solids does not guarantee a similarity in their respective $\chi_{\mathrm{orb}}$.

Interestingly, however, the diamagnetic well depth of $\mathrm{MoS_2}$ is precisely predicted by the square lattice model.
This might lead to the conclusion that the orbital character of the lower and upper band needs to be reflected by the underlying effective tight-binding model in order to describe the magnetic response, because the core gap of $\mathrm{MoS_2}$ is mainly composed of by $p_x$ and $p_y$-orbitals (for lower and upper band) whereas the gap at the neutrality point is made up by $d_{x^2-y^2}$ and $d_{xy}$-orbitals (valence band) as well as predominately $d_{z^2}$-orbitals (conduction band).\cite{Kormanyos13,Cappelluti13,rostami2015valley}
Still, we believe that the same orbital susceptibility obtained from the two models is rather a coincidence and no further conclusions can be drawn.
This is mainly suggested by the Brillouin zone analysis as discussed below.
 
Let us now discuss the contributions to the magnetic response at each $\vec{k}$-point shown in Fig. \ref{FigBHZ} (D).
We choose the Hamiltonian of Eq. (\ref{HBHZ}) with parameters of Eq. (\ref{choice1}), although the other models, in particular the square lattice model, show similar behavior.
The relevant Fermi energies are those either inside or very close to the gap, respective to subplots (a) and (b) of Fig. \ref{FigBHZ} (D).
The corresponding patterns are quite similar to those of Fig. \ref{chikpanelMoS2} (d) and (e).
We thus conclude that the nature of the processes encoding the magnetic response is approximately the same for the two cases.
This reinforces our previous comment about the qualitative agreement between their respective magnetic response.
Concerning the quantitative discrepancies of a model with its effective counterpart, the threefold symmetry of Fig. \ref{chikpanelMoS2} (e) cannot be reproduced by the 2-band models, implying a more complex geometrical contribution to the susceptibility than the product of the Berry curvature and the orbital magnetic moment.\cite{Gao15}

As a consequence of the previous discussion, it seems reasonable to state that the magnetic behavior of a material still remains at reach simply from the knowledge of the band structure.
As for the striking difference between the models discussed in Refs. \onlinecite{raoux2014dia,raoux2015orbital}, we associate it to the presence of a flat band as also argued in Ref. \onlinecite{Gao15}.
Still, an accurate prediction depends on factors like the $k$-dependence of the mass term and the topological regime around the valleys and beyond.
 
\section{Conclusion}
\label{sec_conclusion}

In this paper, we have studied the orbital magnetic susceptibility of graphene and $\mathrm{MoS_2}$ described by effective multi-band tight-binding models.
Like this, contributions from processes around the Fermi surface as well as geometrical aspects involving e.g. the Berry curvature are automatically incorporated.
We obtained new results for the magnetic response for both materials which can be tested experimentally - especially for Fermi energies close to the neutrality point or inside the gap.

More concretely, we calculated $\chi_{\mathrm{orb}}$ for gapless and gapped graphene, dealing with an 8-band Slater-Koster model including also the $\sigma$-orbitals.
This yields an additional $\sim 12\%$ diamagnetic contribution relative to the lattice susceptibility close to half filling, independent of whether the $\pi$-band is gapped or gapless.
This additional contribution to $\chi_{\mathrm{orb}}$ is constant inside the gap around the $\Gamma$-point and of purely geometrical nature.
Still, it is fundamentally different from the geometrical susceptibility associated with the Dirac gap of the $\pi$-bands.

We were further able to identify prominent diamagnetic peaks of $\chi_{\mathrm{orb}}$ with Dirac-cone like band-crossings which are exactly described by the McClure formula.
We expect this delta-like diamagnetic response associated to Dirac cones to be a general geometrical effect due to the infinite Berry curvature, but also the related zero effective mass would give this result.
 
In the case of $\mathrm{MoS_2}$ described by the 12-band model, we have identified two prominent diamagnetic contributions associated to two Dirac-like gaps.
We have shown that the Dirac continuum model is quantitatively sensible only for the one between the core bands.
As for energies close to the neutrality point, our analysis involved the comparison with three different 2-band lattice models, which match the band structure of $\mathrm{MoS_2}$ in the vicinity of the gap but generally yield quantitatively different magnetic susceptibilities.
Interestingly, only the one including two orbitals per site gave an accurate magnitude of the diamagnetism.
The qualitative features of $\chi_{\mathrm{orb}}$ are well reproduced in all cases, though.  

Additionally, we have demonstrated that by analyzing the contribution to the total magnetic response in $\vec{k}$-space, valuable information can be gained to identify the processes.
More concretely, we were able to associate the response either to intraband transitions around the Fermi surface or to geometrical processes around the high-symmetry points $K$ and $M$.
Only the diamagnetic response of the $\sigma$-electrons inside the gap around the $\Gamma$-point could not be attributed to localized interband transitions.
The finding of effective models describing this situation remains to be thoroughly clarified and shall be dealt with in future works.

\begin{acknowledgments}
We acknowledge R. Rold\'an, H. Rostami and V. S\'anchez for useful discussions.
This work has been supported by Spain's MINECO under grants FIS2012-37549-C05-03, FIS2013-48048-P, and FIS2014-57432-P. 
\end{acknowledgments} 

\appendix
\section{Magnetic response of tight-binding models}
We summarize the formalism used to obtain the magnetic response\cite{gomez2011measurable} for arbitrary tight-binding
 models. Particular attention is paid to show its gauge-invariant nature.

\subsection{Hamiltonian and gauge invariance}\label{hamil}
We consider a generic tight-binding Hamiltonian in a 3d lattice
\begin{equation}
H = \sum_{i,j} \; h_{i j} \; \ket{i} \bra{j}\;, 
\end{equation}
where $i (j)$ runs over all orbitals in the lattice.  $ \ket{i}$ is shorthand
 for the state $\ket{\bm r_i,\alpha_i} $, located at the position $\bm r_i $ and
 with orbital index $\alpha_i $, one for each   orbital in the unit
cell. It is convenient to consider this discrete set as part of the usual  continuum $\bracket{\bm r,\alpha}{\bm r',\alpha'} = \delta(\bm r-\bm
r')\;\delta_{\alpha \alpha'}$, where $\ket{\bm r,\alpha}$ is the eigenstate of
the position operator for the orbital index $\alpha $,  $ \bm R_\alpha $, with
 conjugate momentum  $ \bm P_\alpha $. They satisfy  canonical
commutation relations, $ [\hat{\bm n} \cdot \bm R_\alpha , \; \hat{\bm n} \cdot \bm
P_{\alpha'}] = i \, \hbar \, \delta_{\alpha \alpha'}$,  $ \hat{\bm n}$
being an arbitrary unit vector. Notice that they are diagonal in orbital
index. 

In the absence of a magnetic field we have
\begin{equation}\label{disp}
 \ket{\bm r + \bm a,\alpha} =  e^{-\frac{i}{\hbar}\bm a \cdot\bm P} \, \ket{\bm
  r,\alpha}\;, 
\end{equation}
with $\bm P = \sum_{\alpha} \bm P_\alpha $. In the presence of a magnetic field
 with vector potential $\bm {\mathcal A}(\bm r) $ and  operator $\bm A (\bm
  R) =
 \sum_{\alpha} \int d^3r  \bm {\mathcal A}(\bm r) \ket{\bm r,\alpha} 
\bra{\bm r,\alpha} $,  the replacement  $\bm P \rightarrow \bm \Pi = \bm P - e\bm A (\bm
 R)
 $, where $\bm R = \sum_{\alpha} \bm R_\alpha  $, changes Eq. (\ref{disp}) to 
\begin{equation}\label{dispA}
  e^{-\frac{i}{\hbar}\bm a \cdot \bm \Pi}  \, \ket{\bm r,\alpha} = 
 e^{i \phi({\bm r,\bm r + \bm a})} \ket{\bm r + \bm a,\alpha}\;,
\end{equation}
with the Berry's phase for parallel transport becoming here the usual Peierls phase, 
$\phi({\bm r,\bm r'}) = \frac{e}{\hbar}\int_{\bm r}^{\bm r'} d \bm l \cdot \bm {\mathcal A}$. 
 
%
%
 The original Hamiltonian in the absence of the field
%
%
becomes in the presence of the field
\begin{equation}\label{hamilA}
H = \sum_{i,j} \; h_{i j} \; \ket{\bm r_i,\alpha_i} 
\bra{\bm r_i,\alpha_j} \; e^{\frac{i}{\hbar} \bm \delta_{i j} \cdot \bm \Pi}\;,
\end{equation}
with $ \bm \delta_{i j} = \bm r_j - \bm r_i $. This manifestly gauge-invariant form is due to
both  the presence of $ \bm \Pi$ and the shared location of bra and ket in Eq.
 (\ref{hamilA}).

\subsection{Current operator and replicas}\label{Sec:current}
The  previous formulation provides a unique, unambiguous prescription for the current operator anywhere in space. Let us  consider a single oriented hopping term, 
\begin{equation}
 H_{i j} =  h_{i j} \; \ket{\bm r_i,\alpha_i} 
\bra{\bm r_i,\alpha_j} \; e^{\frac{i}{\hbar} \bm \delta_{i j} \cdot \bm \Pi}\;.
\end{equation}
The current operator, given by $ \bm J(\bm r) = -\frac{\delta H}{\delta \bm {\mathcal A}(\bm r)} $,  leads for $H_{i j}$ to   
\begin{align}\label{currentij}
\bm J_{i j}(\bm r) &= \frac{ie}{\hbar}  h_{i j}\, \bm \delta_{i j} 
\ket{\bm r_i,\alpha_i} \bra{\bm r_i,\alpha_j}
\int_0^1 ds \notag\\ 
&\times e^{\frac{i}{\hbar} s \bm  \delta_{i j} \cdot \bm \Pi}
\;\ket{\bm r,\alpha_j} \bra{\bm r,\alpha_j}
\;e^{\frac{i}{\hbar} (1-s) \bm  \delta_{i j} \cdot \bm \Pi}\;,
\end{align}
where the relation $
\delta e^K = \int_0^1 ds \;e^{s K} \;\delta K \;e^{(1-s) K}
$   has been used for dealing with non- commuting operators $K $ and $\delta K$.\cite{stauber2010dynamical}
The point of writing the current in this form is to exhibit its gauge-invariant nature. A more familiar expression would be  
\begin{align}\label{currentijbis}
&
\bm J_{i j}(\bm r) = \frac{ie}{\hbar}  h_{i j}\bm \delta_{i j} 
e^{-i \phi({\bm r_i,\bm r_j})}
\ket{\bm r_i,\alpha_i} \bra{\bm r_j,\alpha_j} \notag
\\
&
\times\int_0^1 ds\delta(\bm r_i - \bm r + s \,\bm \delta_{i j}),
\end{align}
where the last integral fixes the straight line between $\bm r_i $ and $\bm r_j $ as the loci for non zero currents: the familiar network picture now for quantum operators. The continuity equation holds everywhere with source and drain end points. 
 
 The extreme  localization of the network-like current was found inconvenient for the perturbative approach,\cite{gomez2011measurable}  and a continuum of replicas  of the original system obtained by displacing the reference lattice by $\bm \rho $, taken uniformly within the unit cell, was introduced, 
\begin{equation}\label{replicas}
H = \frac{1}{N} \sum_{i,j}h_{i j} \int d^3\rho \ket{\bm r_i + \bm \rho,\alpha_i} 
\bra{\bm r_i + \bm \rho,\alpha_j} e^{\frac{i}{\hbar} \bm \delta_{i j} \cdot \bm \Pi}\;,
\end{equation}
N being the total number of cells. Replicas labeled by $\bm \rho $ are different modulo a lattice vector,  allowing $\bm \rho $ to span all space after  appropriate normalization. Different replicas are dynamically independent: a particle in one of them  will hop in its own discrete lattice, unaware of any of the other replicas,  allowing the average to be taken at the Hamiltonian level. The lattice is displaced but the field is kept in place: each replica experiences a slightly different field, and the process can be interpreted alternatively as an average over slightly displaced fields. This replication will leave properties of the original problem  virtually unaffected, unless the field changes drastically at the lattice length scale, a situation where even the  tight-binding Hamiltonian is questionable. Furthermore,  a translation amounts to  a gauge transformation for a uniform magnetic field,  leaving   physical properties intact. Irrespective of its origin, the manifestly gauge-invariant Hamiltonian of Eq. (\ref{replicas}), leads to the following gauge-invariant current operator, unambiguously  defined everywhere   in space, 
\begin{align}\label{current}
\bm J(\bm r) &= \frac{ie}{\hbar}\frac{1}{N}\sum_i \sum_j \;h_{i j}\, \bm \delta_{i j} \int d^3\rho
\ket{\bm r_i+\bm \rho,\alpha_i} \bra{\bm r_i+\bm \rho,\alpha_j}\notag\\
&\times\int_0^1 ds e^{\frac{i}{\hbar} s \bm  \delta_{i j} \cdot \bm \Pi}
\;\ket{\bm r,\alpha_j} \bra{\bm r,\alpha_j}
\;e^{\frac{i}{\hbar} (1-s) \bm  \delta_{i j} \cdot \bm \Pi}\;.
\end{align}

\subsection{Paramagnetic current, linear response and orbital susceptibility}

In the absence of fields, the Hamiltonian  Bloch matrix,  $\hat{H}_{\bm k} = H_{\alpha \beta}(\bm k) $, is 
\begin{equation}
 H_{\alpha \beta} (\bm k) = \frac{1}{N} \sum_{i(\alpha),j(\beta)} 
 h_{i j} e^{i \bm k \cdot \bm\delta_{i j}}\;,
\end{equation}
where $i(\alpha)$ $(j(\beta))$ runs over all orbitals of $\alpha$ $(\beta)$
 index.  
The paramagnetic current operator in real space reads
\begin{equation}
 \bm J(\bm r) =  \frac{i e}{\hbar}  \frac{1}{N} \sum_{i,j}  
 h_{i j} \bm \delta_{i j} \int_0^1 ds 
 \ket{\bm r - s \bm \delta_{i j},\alpha_i} 
  \bra{\bm r +(1-s) \bm \delta_{i j},\beta_j} 
\end{equation}
with Fourier components
\begin{equation}
 \bm J(\bm q) =  \frac{e}{\hbar} \frac{1}{V^{1/2}} \sum_{\alpha, \beta}
  \sum_{\bm k} \ket{\bm k - \bm q /2,\alpha} 
 \bm \gamma_{\alpha \beta}(\bm k,\bm q)  \bra{\bm k + \bm q /2,\beta}\;,
\end{equation} 
 total volume $V$, and  matrix kernel,  
$ \hat{\bm \gamma}_{\bm k, \bm q} = \bm \gamma_{\alpha \beta}(\bm k,\bm q)$,
     given by
\begin{equation}
\bm \gamma_{\alpha \beta}(\bm k,\bm q) =  \frac{1}{N} \sum_{i(\alpha),j(\beta)} 
  i h_{i j}  \bm \delta_{i j} e^{i \bm k \cdot \bm\delta_{i j}} \text{sinc}
 (\bm q \cdot \bm\delta_{i j}/2)\;, 
\end{equation}
where $\text{sinc}(x) = \frac{\sin(x)}{x}$.
The zero-$\bm q$ limit reads
$\hat{\bm \gamma}_{\bm k}=
\hat{\bm \gamma}_{\bm k, \bm q \rightarrow 0} = 
\nabla_{\bm k} \hat{H}_{\bm k} 
$, so $\langle \bm J(\bm q\rightarrow 0)\rangle $  measures the velocity content of
 Bloch states, as expected. 
 
 In the presence of fields, the Hamiltonian is perturbed to linear order by  $ V = -\sum_{\bm q} \bm J(\bm q) \cdot \bm {\mathcal A}(-\bm q) $, and linear response prescribes the following result for the paramagnetic current
\begin{equation}
\langle\bm J({\bm q})\rangle = \frac{-1}{2 \pi i} \int \!\! dE   n_F(E) 
\text{Tr}\{\bm J({\bm q}) (G^{r} V G^{r} - G^{a} V G^{a})\}\;,
\end{equation}
with retarded and advanced Green function for the unperturbed Hamiltonian, 
$ G^{r,a}(E) = (E \pm i 0^+ -H)^{-1} $, diagonal in Bloch space
$ \hat{G}_{\bm k}^{r,a}(E) = (E \pm i 0^+ - \hat{H}_{\bm k})^{-1} $, leading to the following expression for the paramagnetic response tensor, $\langle  \bm J(\bm q) \rangle =  \bm \chi(\bm q) \; \; \bm {\mathcal A}(\bm q)$,
\begin{align}\label{jtotal}
\bm \chi({\bm q}) = & \frac{e^2}{\hbar^2}\frac{1}{ 2\pi i} 
\int \!\!dE   n_F(E) \frac{1}{V} \times \notag\\
&\sum_{\bm k}
\text{Tr}\{ 
\hat{\bm \gamma}_{\bm k,\bm q}
\hat{ G}^r_{\bm k+\bm q/2}
\hat{\bm \gamma}_{\bm k,-\bm q} 
\hat{ G}^r_{\bm k-\bm q/2}  -
\notag\\
&
\hat{\bm \gamma}_{\bm k,\bm q}
\hat{ G}^a_{\bm k+\bm q/2}
\hat{\bm \gamma}_{\bm k,-\bm q} 
\hat{ G}^a_{\bm k-\bm q/2}
\}\;,
\end{align}
an expression which is valid for arbitrary $\bm q$. 
 To study the low $\bm q$ limit, pertinent for a uniform magnetic field, it is convenient to  define the following auxiliary  tensor, 
\begin{align}
\label{jtotal0}
&\bm \chi_0({\bm q}) = \frac{e^2}{\hbar^2}\frac{1}{ 2\pi i} 
\int \!\!dE n_F(E)\frac{1}{V} \sum_{\bm k}\notag \\
&\times\text{Tr}\{ 
\hat{\bm \gamma}_{\bm k}
\hat{ G}^r_{\bm k+\bm q/2}
\hat{\bm \gamma}_{\bm k} 
\hat{ G}^r_{\bm k-\bm q/2}  - 
\hat{\bm \gamma}_{\bm k}
\hat{ G}^a_{\bm k+\bm q/2}
\hat{\bm \gamma}_{\bm k} 
\hat{ G}^a_{\bm k-\bm q/2}
\}\;, 
\end{align}
where vertex matrices in Eq. (\ref{jtotal}) have been taken at $\bm q = 0 $. The physical response for  a 
 uniform magnetic field, $\bm \chi_{phys}({\bm q})$,
 is given by the $q^2$ term in the expansion of $\bm \chi_0({\bm q}) $: 
\begin{equation}\label{physicalj}
\bm \chi_{phys}({\bm q} \approx 0) = \bm \chi_0({\bm q}) - \bm \chi({\bm q}=0) + 
 \mathcal{O} (q^4)
\;.
\end{equation}
For a uniform magnetic field along an arbitrary direction $\hat{\bm z}$, the orbital magnetic susceptibility corresponds to
\begin{equation}
\frac{\chi_{orb}}{\mu_0}  = \lim_{q\rightarrow 0} \frac{1}{q^2}\chi_{phys}^{y y}(q\hat{\bm x})\;,
\end{equation}
 $x$ and $y$ being orthogonal axis in the plane perpendicular to $\hat{\bm z} $.
 A Taylor expansion of $(\Delta_{\pm} \hat{ G}_{\bm k}) $ to order $q^2$ with repeated use of the relation $\bm \nabla\hat{G}_{\bm k} = \hat{G}_{\bm k} \hat{\bm \gamma}_{\bm k}  \hat{G}_{\bm k} $, and standard manipulations then lead  to  Eq. 1 of the main text. The result, first obtained in Ref. \onlinecite{gomez2011measurable} as a necessary tight-binding generalization  of Fukuyama's result,\cite{fukuyama1971theory} is  reproduced here for completeness:   
\begin{align}\label{chi}
&\chi_{orb} = - \mu_0 \frac{e^2}{\hbar^2}
 \frac{1}{ 2\pi} \text{Im} \int \!\! dE \, n_F(E)\, \frac{1}{V} \sum_{\bm k} \\
 & \text{Tr}\{ 
\hat{\gamma}^{x}
\hat{ G}
\hat{\gamma}^{y}
\hat{ G}
\hat{\gamma}^{x}
\hat{ G}
\hat{\gamma}^{y}
\hat{ G}
+ 
\frac{1}{2}(
\hat{ G}
\hat{\gamma}^{x}
\hat{ G}
\hat{\gamma}^{y}
+
\hat{ G}
\hat{\gamma}^{y}
\hat{ G}
\hat{\gamma}^{x}
)\hat{ G}
\frac{\partial\hat{\gamma}^{y}
 }{\partial k_x}
\}\;, 
\nonumber
\end{align}
with $\bf k$-dependencies removed and $\hat{ G}=\hat{ G}^r$.

There is an additional contribution to the paramagnetic current response,   
$\langle \Delta \bm J(\bm q) \rangle = \Delta \bm \chi(\bm q) \; \; 
\bm {\mathcal A}(\bm q)$,  coming from the ignored $q$-dependence of vertex matrices $\hat{\bm \gamma}_{\bm k,\bm q} $, and 
given to order $q^2$  by the following expression in diadic form 
\begin{equation}\label{deltachi}
\Delta \bm \chi(\bm q) = \frac{e^2}{\hbar^2} \frac{1}{N} \sum_{i,j}
  \langle H_{i j}\rangle   \bm \delta_{i j} \, \bm \delta_{i j} 
  (1 - 2  \text{sinc}(\bm q \cdot \bm\delta_{i j}/2) )\;,
\end{equation}
but it does not show up in  the physical current, 
  being canceled  by the diamagnetic term as we now show. 


\subsection{Diamagnetic current and cancellation}
 Unlike the traditional case, the current operator of Eq. (\ref{current}) has
  terms to all orders in the field. To the linear order relevant here,
   functional differentiation of Eq. (\ref{current})   leads to
    the following expression for the diamagnetic current in real space\cite{stauber2010dynamical}
\begin{align}
\langle \bm J_{dia}(\bm r)\rangle &=  \frac{e^2}{\hbar^2} \frac{1}{N} \sum_{i,j}\bm \delta_{i j} \, \bm
 \delta_{i j}   \langle H_{i j}\rangle  
\int_0^1 ds \int_0^1 ds' \notag\\ &\times[s  \bm {\mathcal A}(\bm r -  s s' \bm \delta_{i j}) +
  s'   \bm {\mathcal A}(\bm r +  s s' \bm \delta_{i j})          ]\;, 
\end{align}
already evaluated in the ground state. In Fourier space,
%
$\langle \bm J_{dia}(\bm q)\rangle = \bm \chi_{dia}(\bm q)
 \bm {\mathcal A}({\bm q}),$  
%
the diamagnetic tensor reads  
\begin{equation}\label{dia}
\bm \chi_{dia}(\bm q) = \frac{e^2}{\hbar^2} \frac{1}{N} \sum_{i,j}
  \langle H_{i j}\rangle   \bm \delta_{i j}   \bm \delta_{i j} 
   \text{sinc}^2(\bm q \cdot \bm\delta_{i j}/2) \;.
\end{equation}  
 In contrast with the usual case, the diamagnetic contribution has $q$-dependence beyond the constant term,\cite{stauber2010dynamical} and its calculation for  a uniform magnetic field has to be completed to order $q^2$. Combining Eq. (\ref{dia}) with the previous contribution from the paramagnetic current, Eq. (\ref{deltachi}), the announced cancellation takes place  
\begin{equation}
\bm \chi_{dia}(\bm q) + \Delta \bm \chi(\bm q) = 0 + \mathcal{O} (q^4)\;.
\end{equation}  
leaving alone   $\bm \chi_{phys} $ as the physical response to a uniform field, with the known expression (\ref{chi}) for the magnetic susceptibility.

\subsection{Absence of longitudinal response}
A longitudinal, static vector potential is a gauge transformation, without
 physical effects. Our gauge invariant  perturbative response should 
   then vanish  to all orders, and  we explicitly 
  show it   to the calculated $q^2$
order. The longitudinal response to a longitudinal vector potential along an arbitrary direction $\hat{\bm x} $ is
  given by 
%
\begin{equation}\label{long}
 \frac{\chi_{phys}^{x x}(q\hat{\bm x})}{q^2/V} \propto  \sum_{\bm k} \\
  \text{Tr}\{ 
\hat{\gamma}^{x}
\hat{ G}
\hat{\gamma}^{x}
\hat{ G}
\hat{\gamma}^{x}
\hat{ G}
\hat{\gamma}^{x}
\hat{ G}
+ 
\hat{ G}
\hat{\gamma}^{x}
\hat{ G}
\hat{\gamma}^{x}
\hat{ G}
\frac{\partial\hat{\gamma}^{x}
 }{\partial k_x}
\} 
,\end{equation}
%
with  factors irrelevant for the argument  ignored. Up to a  total derivative, the second trace cancels the  first one,  
\begin{align}\label{tracelong}
\text{Tr}\{
\hat{ G}
\hat{\gamma}^{x}
\hat{ G}
\hat{\gamma}^{x}
\hat{ G}
\frac{\partial\hat{\gamma}^{x}
 }{\partial k_x}
\} 
 &=  - 
\text{Tr}\{ 
\hat{\gamma}^{x}
\hat{ G}
\hat{\gamma}^{x}
\hat{ G}
\hat{\gamma}^{x}
\hat{ G}
\hat{\gamma}^{x}
\hat{ G}
\} \\
 & + \frac{1}{3}
\frac{\partial}{\partial k_x}
\text{Tr}\{
\hat{ G}
\hat{\gamma}^{x}
\hat{ G}
\hat{\gamma}^{x}
\hat{ G}
\hat{\gamma}^{x}
\}
\nonumber
,\end{align}
and the longitudinal response vanishes, $\chi_{phys}^{x x}(q\hat{\bm x}) = 0$. 
 In a similar way, it can be shown that a longitudinal static perturbation does
   not produce a transverse response, $\chi_{phys}^{y x}(q\hat{\bm x}) = 0$, 
    and vice versa.

  
\subsection{Sum rule}

The susceptibility sum rule, 
\begin{equation}\label{sum}
\int \; d E_F \; \chi_{orb}(E_F)  =  0\;, 
\end{equation}
was ---to the best of our knowledge--- first stated in Ref. \onlinecite{gomez2011measurable}. Its proof from this formalism is direct. Writing as $E_F  \tfrac{d \chi_{orb}}{d E_F} $ the integrand in Eq. (\ref{sum})  from partial integration, then $ \tfrac{d \chi_{orb}}{d E_F}$ is the zero-temperature energy integrand of Eq. (\ref{chi}) evaluated at $E_F$. It is the imaginary part of an  analytic complex function in the upper complex plane, thanks to the presence of products of $\hat{ G}^r$. Closing the
contour with the standard semicircle, where the integral vanishes owing  to the
asymptotic behavior $\hat{ G}^r(z) \sim z^{-1} $, completes the proof. The sum rule  also holds at finite temperatures,  where responses for non-interacting electrons are always a convolution of zero temperature results with the unit area function $\beta/(4 \cosh^2(\beta  \mu/2))$.

\bibliography{mos2_6}

\begin{thebibliography}{30}%
\makeatletter
\providecommand \@ifxundefined [1]{%
 \@ifx{#1\undefined}
}%
\providecommand \@ifnum [1]{%
 \ifnum #1\expandafter \@firstoftwo
 \else \expandafter \@secondoftwo
 \fi
}%
\providecommand \@ifx [1]{%
 \ifx #1\expandafter \@firstoftwo
 \else \expandafter \@secondoftwo
 \fi
}%
\providecommand \natexlab [1]{#1}%
\providecommand \enquote  [1]{``#1''}%
\providecommand \bibnamefont  [1]{#1}%
\providecommand \bibfnamefont [1]{#1}%
\providecommand \citenamefont [1]{#1}%
\providecommand \href@noop [0]{\@secondoftwo}%
\providecommand \href [0]{\begingroup \@sanitize@url \@href}%
\providecommand \@href[1]{\@@startlink{#1}\@@href}%
\providecommand \@@href[1]{\endgroup#1\@@endlink}%
\providecommand \@sanitize@url [0]{\catcode `\\12\catcode `\$12\catcode
  `\&12\catcode `\#12\catcode `\^12\catcode `\_12\catcode `\%12\relax}%
\providecommand \@@startlink[1]{}%
\providecommand \@@endlink[0]{}%
\providecommand \url  [0]{\begingroup\@sanitize@url \@url }%
\providecommand \@url [1]{\endgroup\@href {#1}{\urlprefix }}%
\providecommand \urlprefix  [0]{URL }%
\providecommand \Eprint [0]{\href }%
\providecommand \doibase [0]{http://dx.doi.org/}%
\providecommand \selectlanguage [0]{\@gobble}%
\providecommand \bibinfo  [0]{\@secondoftwo}%
\providecommand \bibfield  [0]{\@secondoftwo}%
\providecommand \translation [1]{[#1]}%
\providecommand \BibitemOpen [0]{}%
\providecommand \bibitemStop [0]{}%
\providecommand \bibitemNoStop [0]{.\EOS\space}%
\providecommand \EOS [0]{\spacefactor3000\relax}%
\providecommand \BibitemShut  [1]{\csname bibitem#1\endcsname}%
\let\auto@bib@innerbib\@empty
\bibitem [{\citenamefont {Novoselov}\ \emph
  {et~al.}(2005{\natexlab{a}})\citenamefont {Novoselov}, \citenamefont {Jiang},
  \citenamefont {Schedin}, \citenamefont {Booth}, \citenamefont {Khotkevich},
  \citenamefont {Morozov},\ and\ \citenamefont {Geim}}]{NovoselovPNAS05}%
  \BibitemOpen
  \bibfield  {author} {\bibinfo {author} {\bibfnamefont {K.~S.}\ \bibnamefont
  {Novoselov}}, \bibinfo {author} {\bibfnamefont {D.}~\bibnamefont {Jiang}},
  \bibinfo {author} {\bibfnamefont {F.}~\bibnamefont {Schedin}}, \bibinfo
  {author} {\bibfnamefont {T.~J.}\ \bibnamefont {Booth}}, \bibinfo {author}
  {\bibfnamefont {V.~V.}\ \bibnamefont {Khotkevich}}, \bibinfo {author}
  {\bibfnamefont {S.~V.}\ \bibnamefont {Morozov}}, \ and\ \bibinfo {author}
  {\bibfnamefont {A.~K.}\ \bibnamefont {Geim}},\ }\href {\doibase
  10.1073/pnas.0502848102} {\bibfield  {journal} {\bibinfo  {journal}
  {Proceedings of the National Academy of Sciences of the United States of
  America}\ }\textbf {\bibinfo {volume} {102}},\ \bibinfo {pages} {10451}
  (\bibinfo {year} {2005}{\natexlab{a}})}\BibitemShut {NoStop}%
\bibitem [{\citenamefont {Blount}(1962)}]{Blount62}%
  \BibitemOpen
  \bibfield  {author} {\bibinfo {author} {\bibfnamefont {E.~I.}\ \bibnamefont
  {Blount}},\ }\href {\doibase 10.1103/PhysRev.126.1636} {\bibfield  {journal}
  {\bibinfo  {journal} {Phys. Rev.}\ }\textbf {\bibinfo {volume} {126}},\
  \bibinfo {pages} {1636} (\bibinfo {year} {1962})}\BibitemShut {NoStop}%
\bibitem [{\citenamefont {Misra}\ and\ \citenamefont {Roth}(1969)}]{Misra69}%
  \BibitemOpen
  \bibfield  {author} {\bibinfo {author} {\bibfnamefont {P.~K.}\ \bibnamefont
  {Misra}}\ and\ \bibinfo {author} {\bibfnamefont {L.~M.}\ \bibnamefont
  {Roth}},\ }\href {\doibase 10.1103/PhysRev.177.1089} {\bibfield  {journal}
  {\bibinfo  {journal} {Phys. Rev.}\ }\textbf {\bibinfo {volume} {177}},\
  \bibinfo {pages} {1089} (\bibinfo {year} {1969})}\BibitemShut {NoStop}%
\bibitem [{\citenamefont {Ashcroft}\ and\ \citenamefont
  {Mermin}(1976)}]{ashcroft1976solid}%
  \BibitemOpen
  \bibfield  {author} {\bibinfo {author} {\bibfnamefont {N.~W.}\ \bibnamefont
  {Ashcroft}}\ and\ \bibinfo {author} {\bibfnamefont {N.~D.}\ \bibnamefont
  {Mermin}},\ }\href@noop {} {\bibfield  {journal} {\bibinfo  {journal}
  {Saunders, Philadelphia}\ ,\ \bibinfo {pages} {293}} (\bibinfo {year}
  {1976})}\BibitemShut {NoStop}%
\bibitem [{\citenamefont {Shi}\ \emph {et~al.}(2007)\citenamefont {Shi},
  \citenamefont {Vignale}, \citenamefont {Xiao},\ and\ \citenamefont
  {Niu}}]{Shi07}%
  \BibitemOpen
  \bibfield  {author} {\bibinfo {author} {\bibfnamefont {J.}~\bibnamefont
  {Shi}}, \bibinfo {author} {\bibfnamefont {G.}~\bibnamefont {Vignale}},
  \bibinfo {author} {\bibfnamefont {D.}~\bibnamefont {Xiao}}, \ and\ \bibinfo
  {author} {\bibfnamefont {Q.}~\bibnamefont {Niu}},\ }\href {\doibase
  10.1103/PhysRevLett.99.197202} {\bibfield  {journal} {\bibinfo  {journal}
  {Phys. Rev. Lett.}\ }\textbf {\bibinfo {volume} {99}},\ \bibinfo {pages}
  {197202} (\bibinfo {year} {2007})}\BibitemShut {NoStop}%
\bibitem [{\citenamefont {Xiao}\ \emph {et~al.}(2010)\citenamefont {Xiao},
  \citenamefont {Chang},\ and\ \citenamefont {Niu}}]{Xiao10}%
  \BibitemOpen
  \bibfield  {author} {\bibinfo {author} {\bibfnamefont {D.}~\bibnamefont
  {Xiao}}, \bibinfo {author} {\bibfnamefont {M.-C.}\ \bibnamefont {Chang}}, \
  and\ \bibinfo {author} {\bibfnamefont {Q.}~\bibnamefont {Niu}},\ }\href
  {\doibase 10.1103/RevModPhys.82.1959} {\bibfield  {journal} {\bibinfo
  {journal} {Rev. Mod. Phys.}\ }\textbf {\bibinfo {volume} {82}},\ \bibinfo
  {pages} {1959} (\bibinfo {year} {2010})}\BibitemShut {NoStop}%
\bibitem [{\citenamefont {Thonhauser}(2011)}]{thonhauser2011theory}%
  \BibitemOpen
  \bibfield  {author} {\bibinfo {author} {\bibfnamefont {T.}~\bibnamefont
  {Thonhauser}},\ }\href@noop {} {\bibfield  {journal} {\bibinfo  {journal}
  {International Journal of Modern Physics B}\ }\textbf {\bibinfo {volume}
  {25}},\ \bibinfo {pages} {1429} (\bibinfo {year} {2011})}\BibitemShut
  {NoStop}%
\bibitem [{\citenamefont {Gao}\ \emph {et~al.}(2014)\citenamefont {Gao},
  \citenamefont {Yang},\ and\ \citenamefont {Niu}}]{Gao14}%
  \BibitemOpen
  \bibfield  {author} {\bibinfo {author} {\bibfnamefont {Y.}~\bibnamefont
  {Gao}}, \bibinfo {author} {\bibfnamefont {S.~A.}\ \bibnamefont {Yang}}, \
  and\ \bibinfo {author} {\bibfnamefont {Q.}~\bibnamefont {Niu}},\ }\href
  {\doibase 10.1103/PhysRevLett.112.166601} {\bibfield  {journal} {\bibinfo
  {journal} {Phys. Rev. Lett.}\ }\textbf {\bibinfo {volume} {112}},\ \bibinfo
  {pages} {166601} (\bibinfo {year} {2014})}\BibitemShut {NoStop}%
\bibitem [{\citenamefont {Raoux}\ \emph {et~al.}(2015)\citenamefont {Raoux},
  \citenamefont {Pi{\'e}chon}, \citenamefont {Fuchs},\ and\ \citenamefont
  {Montambaux}}]{raoux2015orbital}%
  \BibitemOpen
  \bibfield  {author} {\bibinfo {author} {\bibfnamefont {A.}~\bibnamefont
  {Raoux}}, \bibinfo {author} {\bibfnamefont {F.}~\bibnamefont {Pi{\'e}chon}},
  \bibinfo {author} {\bibfnamefont {J.-N.}\ \bibnamefont {Fuchs}}, \ and\
  \bibinfo {author} {\bibfnamefont {G.}~\bibnamefont {Montambaux}},\
  }\href@noop {} {\bibfield  {journal} {\bibinfo  {journal} {Physical Review
  B}\ }\textbf {\bibinfo {volume} {91}},\ \bibinfo {pages} {085120} (\bibinfo
  {year} {2015})}\BibitemShut {NoStop}%
\bibitem [{\citenamefont {Gao}\ \emph {et~al.}(2015)\citenamefont {Gao},
  \citenamefont {Yang},\ and\ \citenamefont {Niu}}]{Gao15}%
  \BibitemOpen
  \bibfield  {author} {\bibinfo {author} {\bibfnamefont {Y.}~\bibnamefont
  {Gao}}, \bibinfo {author} {\bibfnamefont {S.~A.}\ \bibnamefont {Yang}}, \
  and\ \bibinfo {author} {\bibfnamefont {Q.}~\bibnamefont {Niu}},\ }\href
  {\doibase 10.1103/PhysRevB.91.214405} {\bibfield  {journal} {\bibinfo
  {journal} {Phys. Rev. B}\ }\textbf {\bibinfo {volume} {91}},\ \bibinfo
  {pages} {214405} (\bibinfo {year} {2015})}\BibitemShut {NoStop}%
\bibitem [{\citenamefont {Fukuyama}(1971)}]{fukuyama1971theory}%
  \BibitemOpen
  \bibfield  {author} {\bibinfo {author} {\bibfnamefont {H.}~\bibnamefont
  {Fukuyama}},\ }\href@noop {} {\bibfield  {journal} {\bibinfo  {journal}
  {Progress of Theoretical Physics}\ }\textbf {\bibinfo {volume} {45}},\
  \bibinfo {pages} {704} (\bibinfo {year} {1971})}\BibitemShut {NoStop}%
\bibitem [{\citenamefont {Safran}\ and\ \citenamefont
  {DiSalvo}(1979)}]{Safran79}%
  \BibitemOpen
  \bibfield  {author} {\bibinfo {author} {\bibfnamefont {S.~A.}\ \bibnamefont
  {Safran}}\ and\ \bibinfo {author} {\bibfnamefont {F.~J.}\ \bibnamefont
  {DiSalvo}},\ }\href {\doibase 10.1103/PhysRevB.20.4889} {\bibfield  {journal}
  {\bibinfo  {journal} {Phys. Rev. B}\ }\textbf {\bibinfo {volume} {20}},\
  \bibinfo {pages} {4889} (\bibinfo {year} {1979})}\BibitemShut {NoStop}%
\bibitem [{\citenamefont {Koshino}\ and\ \citenamefont
  {Ando}(2007)}]{Koshino07}%
  \BibitemOpen
  \bibfield  {author} {\bibinfo {author} {\bibfnamefont {M.}~\bibnamefont
  {Koshino}}\ and\ \bibinfo {author} {\bibfnamefont {T.}~\bibnamefont {Ando}},\
  }\href {\doibase 10.1103/PhysRevB.76.085425} {\bibfield  {journal} {\bibinfo
  {journal} {Phys. Rev. B}\ }\textbf {\bibinfo {volume} {76}},\ \bibinfo
  {pages} {085425} (\bibinfo {year} {2007})}\BibitemShut {NoStop}%
\bibitem [{\citenamefont {G{\'o}mez-Santos}\ and\ \citenamefont
  {Stauber}(2011)}]{gomez2011measurable}%
  \BibitemOpen
  \bibfield  {author} {\bibinfo {author} {\bibfnamefont {G.}~\bibnamefont
  {G{\'o}mez-Santos}}\ and\ \bibinfo {author} {\bibfnamefont {T.}~\bibnamefont
  {Stauber}},\ }\href@noop {} {\bibfield  {journal} {\bibinfo  {journal}
  {Physical review letters}\ }\textbf {\bibinfo {volume} {106}},\ \bibinfo
  {pages} {045504} (\bibinfo {year} {2011})}\BibitemShut {NoStop}%
\bibitem [{\citenamefont {Vignale}(1991)}]{vignale1991orbital}%
  \BibitemOpen
  \bibfield  {author} {\bibinfo {author} {\bibfnamefont {G.}~\bibnamefont
  {Vignale}},\ }\href@noop {} {\bibfield  {journal} {\bibinfo  {journal}
  {Physical review letters}\ }\textbf {\bibinfo {volume} {67}},\ \bibinfo
  {pages} {358} (\bibinfo {year} {1991})}\BibitemShut {NoStop}%
\bibitem [{\citenamefont {Koshino}\ and\ \citenamefont
  {Ando}(2010)}]{koshino2010anomalous}%
  \BibitemOpen
  \bibfield  {author} {\bibinfo {author} {\bibfnamefont {M.}~\bibnamefont
  {Koshino}}\ and\ \bibinfo {author} {\bibfnamefont {T.}~\bibnamefont {Ando}},\
  }\href@noop {} {\bibfield  {journal} {\bibinfo  {journal} {Physical Review
  B}\ }\textbf {\bibinfo {volume} {81}},\ \bibinfo {pages} {195431} (\bibinfo
  {year} {2010})}\BibitemShut {NoStop}%
\bibitem [{\citenamefont {Peierls}(1933)}]{peierls1933theory}%
  \BibitemOpen
  \bibfield  {author} {\bibinfo {author} {\bibfnamefont {R.}~\bibnamefont
  {Peierls}},\ }\href@noop {} {\bibfield  {journal} {\bibinfo  {journal} {Z.
  Phys}\ }\textbf {\bibinfo {volume} {80}},\ \bibinfo {pages} {763} (\bibinfo
  {year} {1933})}\BibitemShut {NoStop}%
\bibitem [{\citenamefont {Ominato}\ and\ \citenamefont
  {Koshino}(2013)}]{Ominato13}%
  \BibitemOpen
  \bibfield  {author} {\bibinfo {author} {\bibfnamefont {Y.}~\bibnamefont
  {Ominato}}\ and\ \bibinfo {author} {\bibfnamefont {M.}~\bibnamefont
  {Koshino}},\ }\href {\doibase http://dx.doi.org/10.1016/j.ssc.2013.09.023}
  {\bibfield  {journal} {\bibinfo  {journal} {Solid State Communications}\
  }\textbf {\bibinfo {volume} {175–176}},\ \bibinfo {pages} {51 } (\bibinfo
  {year} {2013})},\ \bibinfo {note} {special Issue: Graphene V: Recent Advances
  in Studies of Graphene and Graphene analogues}\BibitemShut {NoStop}%
\bibitem [{\citenamefont {Raoux}\ \emph {et~al.}(2014)\citenamefont {Raoux},
  \citenamefont {Morigi}, \citenamefont {Fuchs}, \citenamefont {Pi{\'e}chon},\
  and\ \citenamefont {Montambaux}}]{raoux2014dia}%
  \BibitemOpen
  \bibfield  {author} {\bibinfo {author} {\bibfnamefont {A.}~\bibnamefont
  {Raoux}}, \bibinfo {author} {\bibfnamefont {M.}~\bibnamefont {Morigi}},
  \bibinfo {author} {\bibfnamefont {J.-N.}\ \bibnamefont {Fuchs}}, \bibinfo
  {author} {\bibfnamefont {F.}~\bibnamefont {Pi{\'e}chon}}, \ and\ \bibinfo
  {author} {\bibfnamefont {G.}~\bibnamefont {Montambaux}},\ }\href@noop {}
  {\bibfield  {journal} {\bibinfo  {journal} {Physical review letters}\
  }\textbf {\bibinfo {volume} {112}},\ \bibinfo {pages} {026402} (\bibinfo
  {year} {2014})}\BibitemShut {NoStop}%
\bibitem [{\citenamefont {Novoselov}\ \emph
  {et~al.}(2005{\natexlab{b}})\citenamefont {Novoselov}, \citenamefont {Geim},
  \citenamefont {Morozov}, \citenamefont {Jiang}, \citenamefont {Katsnelson},
  \citenamefont {Grigorieva}, \citenamefont {Dubonos},\ and\ \citenamefont
  {Firsov}}]{Novoselov05}%
  \BibitemOpen
  \bibfield  {author} {\bibinfo {author} {\bibfnamefont {K.~S.}\ \bibnamefont
  {Novoselov}}, \bibinfo {author} {\bibfnamefont {A.~K.}\ \bibnamefont {Geim}},
  \bibinfo {author} {\bibfnamefont {S.~V.}\ \bibnamefont {Morozov}}, \bibinfo
  {author} {\bibfnamefont {D.}~\bibnamefont {Jiang}}, \bibinfo {author}
  {\bibfnamefont {M.~I.}\ \bibnamefont {Katsnelson}}, \bibinfo {author}
  {\bibfnamefont {I.~V.}\ \bibnamefont {Grigorieva}}, \bibinfo {author}
  {\bibfnamefont {S.~V.}\ \bibnamefont {Dubonos}}, \ and\ \bibinfo {author}
  {\bibfnamefont {A.~A.}\ \bibnamefont {Firsov}},\ }\href
  {http://dx.doi.org/10.1038/nature04233} {\bibfield  {journal} {\bibinfo
  {journal} {Nature}\ }\textbf {\bibinfo {volume} {438}},\ \bibinfo {pages}
  {197} (\bibinfo {year} {2005}{\natexlab{b}})}\BibitemShut {NoStop}%
\bibitem [{\citenamefont {Zhang}\ \emph {et~al.}(2005)\citenamefont {Zhang},
  \citenamefont {Tan}, \citenamefont {Stormer},\ and\ \citenamefont
  {Kim}}]{Zhang05}%
  \BibitemOpen
  \bibfield  {author} {\bibinfo {author} {\bibfnamefont {Y.}~\bibnamefont
  {Zhang}}, \bibinfo {author} {\bibfnamefont {Y.-W.}\ \bibnamefont {Tan}},
  \bibinfo {author} {\bibfnamefont {H.~L.}\ \bibnamefont {Stormer}}, \ and\
  \bibinfo {author} {\bibfnamefont {P.}~\bibnamefont {Kim}},\ }\href
  {http://dx.doi.org/10.1038/nature04235} {\bibfield  {journal} {\bibinfo
  {journal} {Nature}\ }\textbf {\bibinfo {volume} {438}},\ \bibinfo {pages}
  {201} (\bibinfo {year} {2005})}\BibitemShut {NoStop}%
\bibitem [{\citenamefont {Gorbachev}\ \emph {et~al.}(2014)\citenamefont
  {Gorbachev}, \citenamefont {Song}, \citenamefont {Yu}, \citenamefont
  {Kretinin}, \citenamefont {Withers}, \citenamefont {Cao}, \citenamefont
  {Mishchenko}, \citenamefont {Grigorieva}, \citenamefont {Novoselov},
  \citenamefont {Levitov},\ and\ \citenamefont {Geim}}]{Gorbachev14}%
  \BibitemOpen
  \bibfield  {author} {\bibinfo {author} {\bibfnamefont {R.~V.}\ \bibnamefont
  {Gorbachev}}, \bibinfo {author} {\bibfnamefont {J.~C.~W.}\ \bibnamefont
  {Song}}, \bibinfo {author} {\bibfnamefont {G.~L.}\ \bibnamefont {Yu}},
  \bibinfo {author} {\bibfnamefont {A.~V.}\ \bibnamefont {Kretinin}}, \bibinfo
  {author} {\bibfnamefont {F.}~\bibnamefont {Withers}}, \bibinfo {author}
  {\bibfnamefont {Y.}~\bibnamefont {Cao}}, \bibinfo {author} {\bibfnamefont
  {A.}~\bibnamefont {Mishchenko}}, \bibinfo {author} {\bibfnamefont {I.~V.}\
  \bibnamefont {Grigorieva}}, \bibinfo {author} {\bibfnamefont {K.~S.}\
  \bibnamefont {Novoselov}}, \bibinfo {author} {\bibfnamefont {L.~S.}\
  \bibnamefont {Levitov}}, \ and\ \bibinfo {author} {\bibfnamefont {A.~K.}\
  \bibnamefont {Geim}},\ }\href {\doibase 10.1126/science.1254966} {\bibfield
  {journal} {\bibinfo  {journal} {Science}\ }\textbf {\bibinfo {volume}
  {346}},\ \bibinfo {pages} {448} (\bibinfo {year} {2014})}\BibitemShut
  {NoStop}%
\bibitem [{\citenamefont {Feng}\ \emph {et~al.}(2012)\citenamefont {Feng},
  \citenamefont {Yao}, \citenamefont {Zhu}, \citenamefont {Zhou}, \citenamefont
  {Yao},\ and\ \citenamefont {Xiao}}]{Feng12}%
  \BibitemOpen
  \bibfield  {author} {\bibinfo {author} {\bibfnamefont {W.}~\bibnamefont
  {Feng}}, \bibinfo {author} {\bibfnamefont {Y.}~\bibnamefont {Yao}}, \bibinfo
  {author} {\bibfnamefont {W.}~\bibnamefont {Zhu}}, \bibinfo {author}
  {\bibfnamefont {J.}~\bibnamefont {Zhou}}, \bibinfo {author} {\bibfnamefont
  {W.}~\bibnamefont {Yao}}, \ and\ \bibinfo {author} {\bibfnamefont
  {D.}~\bibnamefont {Xiao}},\ }\href {\doibase 10.1103/PhysRevB.86.165108}
  {\bibfield  {journal} {\bibinfo  {journal} {Phys. Rev. B}\ }\textbf {\bibinfo
  {volume} {86}},\ \bibinfo {pages} {165108} (\bibinfo {year}
  {2012})}\BibitemShut {NoStop}%
\bibitem [{\citenamefont {Rostami}\ and\ \citenamefont
  {Asgari}(2015)}]{rostami2015valley}%
  \BibitemOpen
  \bibfield  {author} {\bibinfo {author} {\bibfnamefont {H.}~\bibnamefont
  {Rostami}}\ and\ \bibinfo {author} {\bibfnamefont {R.}~\bibnamefont
  {Asgari}},\ }\href@noop {} {\bibfield  {journal} {\bibinfo  {journal}
  {Physical Review B}\ }\textbf {\bibinfo {volume} {91}},\ \bibinfo {pages}
  {075433} (\bibinfo {year} {2015})}\BibitemShut {NoStop}%
\bibitem [{\citenamefont {Korm\'anyos}\ \emph {et~al.}(2013)\citenamefont
  {Korm\'anyos}, \citenamefont {Z\'olyomi}, \citenamefont {Drummond},
  \citenamefont {Rakyta}, \citenamefont {Burkard},\ and\ \citenamefont
  {Fal'ko}}]{Kormanyos13}%
  \BibitemOpen
  \bibfield  {author} {\bibinfo {author} {\bibfnamefont {A.}~\bibnamefont
  {Korm\'anyos}}, \bibinfo {author} {\bibfnamefont {V.}~\bibnamefont
  {Z\'olyomi}}, \bibinfo {author} {\bibfnamefont {N.~D.}\ \bibnamefont
  {Drummond}}, \bibinfo {author} {\bibfnamefont {P.}~\bibnamefont {Rakyta}},
  \bibinfo {author} {\bibfnamefont {G.}~\bibnamefont {Burkard}}, \ and\
  \bibinfo {author} {\bibfnamefont {V.~I.}\ \bibnamefont {Fal'ko}},\ }\href
  {\doibase 10.1103/PhysRevB.88.045416} {\bibfield  {journal} {\bibinfo
  {journal} {Phys. Rev. B}\ }\textbf {\bibinfo {volume} {88}},\ \bibinfo
  {pages} {045416} (\bibinfo {year} {2013})}\BibitemShut {NoStop}%
\bibitem [{\citenamefont {Cappelluti}\ \emph {et~al.}(2013)\citenamefont
  {Cappelluti}, \citenamefont {Rold\'an}, \citenamefont {Silva-Guill\'en},
  \citenamefont {Ordej\'on},\ and\ \citenamefont {Guinea}}]{Cappelluti13}%
  \BibitemOpen
  \bibfield  {author} {\bibinfo {author} {\bibfnamefont {E.}~\bibnamefont
  {Cappelluti}}, \bibinfo {author} {\bibfnamefont {R.}~\bibnamefont
  {Rold\'an}}, \bibinfo {author} {\bibfnamefont {J.}~\bibnamefont
  {Silva-Guill\'en}}, \bibinfo {author} {\bibfnamefont {P.}~\bibnamefont
  {Ordej\'on}}, \ and\ \bibinfo {author} {\bibfnamefont {F.}~\bibnamefont
  {Guinea}},\ }\href {\doibase 10.1103/PhysRevB.88.075409} {\bibfield
  {journal} {\bibinfo  {journal} {Phys. Rev. B}\ }\textbf {\bibinfo {volume}
  {88}},\ \bibinfo {pages} {075409} (\bibinfo {year} {2013})}\BibitemShut
  {NoStop}%
\bibitem [{\citenamefont {Stauber}\ and\ \citenamefont
  {G{\'o}mez-Santos}(2010)}]{stauber2010dynamical}%
  \BibitemOpen
  \bibfield  {author} {\bibinfo {author} {\bibfnamefont {T.}~\bibnamefont
  {Stauber}}\ and\ \bibinfo {author} {\bibfnamefont {G.}~\bibnamefont
  {G{\'o}mez-Santos}},\ }\href@noop {} {\bibfield  {journal} {\bibinfo
  {journal} {Physical Review B}\ }\textbf {\bibinfo {volume} {82}},\ \bibinfo
  {pages} {155412} (\bibinfo {year} {2010})}\BibitemShut {NoStop}%
\bibitem [{\citenamefont {McClure}(1956)}]{McClure56}%
  \BibitemOpen
  \bibfield  {author} {\bibinfo {author} {\bibfnamefont {J.~W.}\ \bibnamefont
  {McClure}},\ }\href {\doibase 10.1103/PhysRev.104.666} {\bibfield  {journal}
  {\bibinfo  {journal} {Phys. Rev.}\ }\textbf {\bibinfo {volume} {104}},\
  \bibinfo {pages} {666} (\bibinfo {year} {1956})}\BibitemShut {NoStop}%
\bibitem [{\citenamefont {McClure}(1960)}]{McClure60}%
  \BibitemOpen
  \bibfield  {author} {\bibinfo {author} {\bibfnamefont {J.~W.}\ \bibnamefont
  {McClure}},\ }\href {\doibase 10.1103/PhysRev.119.606} {\bibfield  {journal}
  {\bibinfo  {journal} {Phys. Rev.}\ }\textbf {\bibinfo {volume} {119}},\
  \bibinfo {pages} {606} (\bibinfo {year} {1960})}\BibitemShut {NoStop}%
\bibitem [{\citenamefont {Yuan}\ \emph {et~al.}(2015)\citenamefont {Yuan},
  \citenamefont {R\"osner}, \citenamefont {Schulz}, \citenamefont {Wehling},\
  and\ \citenamefont {Katsnelson}}]{shengjun2015electronic}%
  \BibitemOpen
  \bibfield  {author} {\bibinfo {author} {\bibfnamefont {S.}~\bibnamefont
  {Yuan}}, \bibinfo {author} {\bibfnamefont {M.}~\bibnamefont {R\"osner}},
  \bibinfo {author} {\bibfnamefont {A.}~\bibnamefont {Schulz}}, \bibinfo
  {author} {\bibfnamefont {T.~O.}\ \bibnamefont {Wehling}}, \ and\ \bibinfo
  {author} {\bibfnamefont {M.~I.}\ \bibnamefont {Katsnelson}},\ }\href
  {\doibase 10.1103/PhysRevLett.114.047403} {\bibfield  {journal} {\bibinfo
  {journal} {Phys. Rev. Lett.}\ }\textbf {\bibinfo {volume} {114}},\ \bibinfo
  {pages} {047403} (\bibinfo {year} {2015})}\BibitemShut {NoStop}%
\end{thebibliography}%

\end{document}